\documentclass[manuscript]{emulateapj}

\usepackage{graphicx}
\usepackage{dcolumn}
\usepackage{bm}
\usepackage{xcolor}
\usepackage{color}
\usepackage{verbatim}
\def\samples{100~}

\def\gammabias{$0.004~$}

\def\HIZcbmean{$-0.079 \pm 0.026$}
\def\HIZcbstdl{$0.021 \pm 0.016$}
\def\HIZcbstdr{$0.098 \pm 0.025$}

\def\HIZccmean{$-0.082 \pm 0.02$}
\def\HIZccstdl{$0.016 \pm 0.013$}
\def\HIZccstdr{$0.098 \pm 0.018$}

\def\HIZcdmean{$-0.085 \pm 0.015$}
\def\HIZcdstdl{$0.014 \pm 0.01$}
\def\HIZcdstdr{$0.088 \pm 0.014$}

\def\HIZcemean{$-0.077 \pm 0.023$}
\def\HIZcestdl{$0.027 \pm 0.015$}
\def\HIZcestdr{$0.1 \pm 0.017$}

\def\HIZcfmean{$-0.084 \pm 0.019$}
\def\HIZcfstdl{$0.022 \pm 0.013$}
\def\HIZcfstdr{$0.108 \pm 0.015$}

\def\HIZcgmean{$-0.081 \pm 0.019$}
\def\HIZcgstdl{$0.024 \pm 0.013$}
\def\HIZcgstdr{$0.11 \pm 0.015$}

\def\HIZchmean{$-0.074 \pm 0.02$}
\def\HIZchstdl{$0.028 \pm 0.013$}
\def\HIZchstdr{$0.111 \pm 0.015$}

\def\HIZcimean{$-0.063 \pm 0.019$}
\def\HIZcistdl{$0.034 \pm 0.013$}
\def\HIZcistdr{$0.114 \pm 0.016$}

\def\HIZcjmean{$-0.062 \pm 0.02$}
\def\HIZcjstdl{$0.034 \pm 0.014$}
\def\HIZcjstdr{$0.117 \pm 0.016$}

\def\HIZckmean{$-0.086 \pm 0.012$}
\def\HIZckstdl{$0.012 \pm 0.009$}
\def\HIZckstdr{$0.138 \pm 0.012$}

\def\HIZclmean{$-0.08 \pm 0.015$}
\def\HIZclstdl{$0.018 \pm 0.011$}
\def\HIZclstdr{$0.138 \pm 0.013$}

\def\HIZcmmean{$-0.079 \pm 0.016$}
\def\HIZcmstdl{$0.019 \pm 0.011$}
\def\HIZcmstdr{$0.139 \pm 0.013$}

\def\HIZcnmean{$-0.063 \pm 0.018$}
\def\HIZcnstdl{$0.03 \pm 0.013$}
\def\HIZcnstdr{$0.125 \pm 0.014$}

\def\HIZcomean{$-0.064 \pm 0.018$}
\def\HIZcostdl{$0.033 \pm 0.012$}
\def\HIZcostdr{$0.125 \pm 0.014$}

\def\HIZcpmean{$-0.068 \pm 0.018$}
\def\HIZcpstdl{$0.031 \pm 0.013$}
\def\HIZcpstdr{$0.128 \pm 0.015$}

\def\HIZcqmean{$-0.062 \pm 0.019$}
\def\HIZcqstdl{$0.036 \pm 0.013$}
\def\HIZcqstdr{$0.123 \pm 0.015$}

\def\HIZcrmean{$-0.065 \pm 0.022$}
\def\HIZcrstdl{$0.039 \pm 0.016$}
\def\HIZcrstdr{$0.125 \pm 0.019$}

\def\HIZcsmean{$-0.074 \pm 0.025$}
\def\HIZcsstdl{$0.034 \pm 0.017$}
\def\HIZcsstdr{$0.118 \pm 0.023$}

\def\HIZctmean{$-0.098 \pm 0.023$}
\def\HIZctstdl{$0.022 \pm 0.016$}
\def\HIZctstdr{$0.138 \pm 0.029$}

\def\HIZcumean{$-0.073 \pm 0.031$}
\def\HIZcustdl{$0.03 \pm 0.023$}
\def\HIZcustdr{$0.118 \pm 0.049$}

\def\HIZxbmean{$0.908 \pm 0.288$}
\def\HIZxbstdl{$0.8 \pm 0.241$}
\def\HIZxbstdr{$0.249 \pm 0.201$}

\def\HIZxcmean{$0.39 \pm 0.349$}
\def\HIZxcstdl{$0.377 \pm 0.245$}
\def\HIZxcstdr{$0.631 \pm 0.244$}

\def\HIZxdmean{$0.511 \pm 0.416$}
\def\HIZxdstdl{$0.633 \pm 0.305$}
\def\HIZxdstdr{$0.575 \pm 0.287$}

\def\HIZxemean{$0.537 \pm 0.365$}
\def\HIZxestdl{$0.571 \pm 0.265$}
\def\HIZxestdr{$0.464 \pm 0.244$}

\def\HIZxfmean{$0.232 \pm 0.277$}
\def\HIZxfstdl{$0.374 \pm 0.202$}
\def\HIZxfstdr{$0.688 \pm 0.188$}

\def\HIZxgmean{$0.242 \pm 0.253$}
\def\HIZxgstdl{$0.421 \pm 0.176$}
\def\HIZxgstdr{$0.685 \pm 0.171$}

\def\HIZxhmean{$0.173 \pm 0.273$}
\def\HIZxhstdl{$0.43 \pm 0.206$}
\def\HIZxhstdr{$0.761 \pm 0.172$}

\def\HIZximean{$0.43 \pm 0.31$}
\def\HIZxistdl{$0.681 \pm 0.228$}
\def\HIZxistdr{$0.591 \pm 0.21$}

\def\HIZxjmean{$0.761 \pm 0.257$}
\def\HIZxjstdl{$1.023 \pm 0.177$}
\def\HIZxjstdr{$0.336 \pm 0.192$}

\def\HIZxkmean{$0.871 \pm 0.225$}
\def\HIZxkstdl{$1.198 \pm 0.155$}
\def\HIZxkstdr{$0.274 \pm 0.169$}

\def\HIZxlmean{$0.938 \pm 0.202$}
\def\HIZxlstdl{$1.422 \pm 0.142$}
\def\HIZxlstdr{$0.218 \pm 0.15$}

\def\HIZxmmean{$0.793 \pm 0.26$}
\def\HIZxmstdl{$1.553 \pm 0.187$}
\def\HIZxmstdr{$0.323 \pm 0.189$}

\def\HIZxnmean{$0.868 \pm 0.221$}
\def\HIZxnstdl{$1.741 \pm 0.162$}
\def\HIZxnstdr{$0.222 \pm 0.149$}

\def\HIZxomean{$0.697 \pm 0.26$}
\def\HIZxostdl{$1.763 \pm 0.186$}
\def\HIZxostdr{$0.33 \pm 0.176$}

\def\HIZxpmean{$0.655 \pm 0.285$}
\def\HIZxpstdl{$1.836 \pm 0.207$}
\def\HIZxpstdr{$0.353 \pm 0.191$}

\def\HIZxqmean{$0.27 \pm 0.334$}
\def\HIZxqstdl{$1.614 \pm 0.25$}
\def\HIZxqstdr{$0.612 \pm 0.227$}

\def\HIZxrmean{$0.062 \pm 0.41$}
\def\HIZxrstdl{$1.571 \pm 0.301$}
\def\HIZxrstdr{$0.745 \pm 0.281$}

\def\HIZxsmean{$0.079 \pm 0.512$}
\def\HIZxsstdl{$1.433 \pm 0.398$}
\def\HIZxsstdr{$0.681 \pm 0.325$}

\def\HIZxtmean{$0.309 \pm 0.512$}
\def\HIZxtstdl{$1.852 \pm 0.428$}
\def\HIZxtstdr{$0.535 \pm 0.363$}

\def\HIZxumean{$-0.14 \pm 0.845$}
\def\HIZxustdl{$1.947 \pm 0.69$}
\def\HIZxustdr{$1.137 \pm 0.672$}

\def\HIZxCbmean{$0.559 \pm 0.227$}
\def\HIZxCbstdl{$0.726 \pm 0.751$}
\def\HIZxCbstdr{$0.729 \pm 0.687$}

\def\HIZxCcmean{$0.473 \pm 0.312$}
\def\HIZxCcstdl{$0.584 \pm 0.611$}
\def\HIZxCcstdr{$0.725 \pm 0.573$}

\def\HIZxCdmean{$0.832 \pm 0.281$}
\def\HIZxCdstdl{$0.835 \pm 0.212$}
\def\HIZxCdstdr{$0.307 \pm 0.19$}

\def\HIZxCemean{$0.791 \pm 0.253$}
\def\HIZxCestdl{$0.796 \pm 0.182$}
\def\HIZxCestdr{$0.288 \pm 0.175$}

\def\HIZxCfmean{$0.458 \pm 0.273$}
\def\HIZxCfstdl{$0.575 \pm 0.196$}
\def\HIZxCfstdr{$0.532 \pm 0.19$}

\def\HIZxCgmean{$0.375 \pm 0.287$}
\def\HIZxCgstdl{$0.543 \pm 0.201$}
\def\HIZxCgstdr{$0.625 \pm 0.198$}

\def\HIZxChmean{$0.59 \pm 0.315$}
\def\HIZxChstdl{$0.766 \pm 0.229$}
\def\HIZxChstdr{$0.517 \pm 0.214$}

\def\HIZxCimean{$0.52 \pm 0.295$}
\def\HIZxCistdl{$0.756 \pm 0.21$}
\def\HIZxCistdr{$0.54 \pm 0.201$}

\def\HIZxCjmean{$0.608 \pm 0.246$}
\def\HIZxCjstdl{$0.892 \pm 0.171$}
\def\HIZxCjstdr{$0.48 \pm 0.178$}

\def\HIZxCkmean{$0.763 \pm 0.239$}
\def\HIZxCkstdl{$1.088 \pm 0.166$}
\def\HIZxCkstdr{$0.352 \pm 0.174$}

\def\HIZxClmean{$0.757 \pm 0.262$}
\def\HIZxClstdl{$1.272 \pm 0.179$}
\def\HIZxClstdr{$0.352 \pm 0.193$}

\def\HIZxCmmean{$0.441 \pm 0.307$}
\def\HIZxCmstdl{$1.138 \pm 0.212$}
\def\HIZxCmstdr{$0.594 \pm 0.22$}

\def\HIZxCnmean{$0.67 \pm 0.309$}
\def\HIZxCnstdl{$1.429 \pm 0.227$}
\def\HIZxCnstdr{$0.389 \pm 0.218$}

\def\HIZxComean{$0.597 \pm 0.338$}
\def\HIZxCostdl{$1.517 \pm 0.246$}
\def\HIZxCostdr{$0.433 \pm 0.237$}

\def\HIZxCpmean{$0.585 \pm 0.35$}
\def\HIZxCpstdl{$1.579 \pm 0.258$}
\def\HIZxCpstdr{$0.431 \pm 0.243$}

\def\HIZxCqmean{$0.513 \pm 0.344$}
\def\HIZxCqstdl{$1.616 \pm 0.257$}
\def\HIZxCqstdr{$0.463 \pm 0.241$}

\def\HIZxCrmean{$0.73 \pm 0.243$}
\def\HIZxCrstdl{$1.79 \pm 0.192$}
\def\HIZxCrstdr{$0.253 \pm 0.17$}

\def\HIZxCsmean{$0.637 \pm 0.31$}
\def\HIZxCsstdl{$1.671 \pm 0.258$}
\def\HIZxCsstdr{$0.323 \pm 0.204$}

\def\HIZxCtmean{$0.608 \pm 0.392$}
\def\HIZxCtstdl{$1.828 \pm 0.547$}
\def\HIZxCtstdr{$0.352 \pm 0.274$}

\def\HIZxCumean{$0.379 \pm 0.605$}
\def\HIZxCustdl{$2.12 \pm 0.538$}
\def\HIZxCustdr{$0.793 \pm 0.657$}

\def\HIZcCbmean{$-0.07 \pm 0.029$}
\def\HIZcCbstdl{$0.023 \pm 0.019$}
\def\HIZcCbstdr{$0.085 \pm 0.027$}

\def\HIZcCcmean{$-0.048 \pm 0.03$}
\def\HIZcCcstdl{$0.024 \pm 0.018$}
\def\HIZcCcstdr{$0.062 \pm 0.023$}

\def\HIZcCdmean{$-0.057 \pm 0.021$}
\def\HIZcCdstdl{$0.016 \pm 0.012$}
\def\HIZcCdstdr{$0.047 \pm 0.02$}

\def\HIZcCemean{$-0.072 \pm 0.022$}
\def\HIZcCestdl{$0.018 \pm 0.014$}
\def\HIZcCestdr{$0.068 \pm 0.018$}

\def\HIZcCfmean{$-0.077 \pm 0.018$}
\def\HIZcCfstdl{$0.014 \pm 0.011$}
\def\HIZcCfstdr{$0.076 \pm 0.015$}

\def\HIZcCgmean{$-0.054 \pm 0.025$}
\def\HIZcCgstdl{$0.024 \pm 0.015$}
\def\HIZcCgstdr{$0.049 \pm 0.018$}

\def\HIZcChmean{$-0.065 \pm 0.021$}
\def\HIZcChstdl{$0.02 \pm 0.013$}
\def\HIZcChstdr{$0.063 \pm 0.015$}

\def\HIZcCimean{$-0.06 \pm 0.02$}
\def\HIZcCistdl{$0.02 \pm 0.013$}
\def\HIZcCistdr{$0.06 \pm 0.015$}

\def\HIZcCjmean{$-0.034 \pm 0.026$}
\def\HIZcCjstdl{$0.033 \pm 0.017$}
\def\HIZcCjstdr{$0.046 \pm 0.018$}

\def\HIZcCkmean{$-0.063 \pm 0.017$}
\def\HIZcCkstdl{$0.017 \pm 0.012$}
\def\HIZcCkstdr{$0.075 \pm 0.012$}

\def\HIZcClmean{$-0.073 \pm 0.014$}
\def\HIZcClstdl{$0.014 \pm 0.009$}
\def\HIZcClstdr{$0.096 \pm 0.011$}

\def\HIZcCmmean{$-0.079 \pm 0.015$}
\def\HIZcCmstdl{$0.014 \pm 0.01$}
\def\HIZcCmstdr{$0.104 \pm 0.011$}

\def\HIZcCnmean{$-0.073 \pm 0.016$}
\def\HIZcCnstdl{$0.017 \pm 0.011$}
\def\HIZcCnstdr{$0.098 \pm 0.012$}

\def\HIZcComean{$-0.079 \pm 0.017$}
\def\HIZcCostdl{$0.018 \pm 0.013$}
\def\HIZcCostdr{$0.106 \pm 0.011$}

\def\HIZcCpmean{$-0.085 \pm 0.014$}
\def\HIZcCpstdl{$0.014 \pm 0.011$}
\def\HIZcCpstdr{$0.11 \pm 0.01$}

\def\HIZcCqmean{$-0.085 \pm 0.015$}
\def\HIZcCqstdl{$0.015 \pm 0.011$}
\def\HIZcCqstdr{$0.108 \pm 0.011$}

\def\HIZcCrmean{$-0.086 \pm 0.019$}
\def\HIZcCrstdl{$0.018 \pm 0.013$}
\def\HIZcCrstdr{$0.109 \pm 0.016$}

\def\HIZcCsmean{$-0.084 \pm 0.021$}
\def\HIZcCsstdl{$0.02 \pm 0.014$}
\def\HIZcCsstdr{$0.097 \pm 0.018$}

\def\HIZcCtmean{$-0.097 \pm 0.026$}
\def\HIZcCtstdl{$0.022 \pm 0.017$}
\def\HIZcCtstdr{$0.111 \pm 0.026$}

\def\HIZcCumean{$-0.06 \pm 0.04$}
\def\HIZcCustdl{$0.036 \pm 0.028$}
\def\HIZcCustdr{$0.091 \pm 0.047$}

\def\ddfst{$0.138 \pm 0.005$}
\def\bbqst{$0.151 \pm 0.005$}
\def\eeast{$0.116 \pm 0.005$}
\def\ddast{$0.149 \pm 0.005$}
\def\ddqst{$0.149 \pm 0.005$}
\def\bbcst{$0.131 \pm 0.005$}
\def\bbast{$0.151 \pm 0.005$}
\def\ddcst{$0.137 \pm 0.005$}
\def\eeqst{$0.121 \pm 0.005$}
\def\eenst{$0.104 \pm 0.005$}
\def\bbfst{$0.132 \pm 0.005$}

\def\bbfa{$0.1444 \pm 0.0005$}
\def\bbfb{$3.7740 \pm 0.0073$}
\def\bbfg{$0.0287 \pm 0.0009$}
\def\bbqa{$0.1458 \pm 0.0005$}
\def\bbqb{$3.2285 \pm 0.0068$}
\def\bbqg{$0.0445 \pm 0.0011$}
\def\bbaa{$0.1459 \pm 0.0005$}
\def\bbab{$3.2283 \pm 0.0068$}
\def\bbag{$0.0432 \pm 0.0011$}
\def\bbca{$0.1460 \pm 0.0005$}
\def\bbcb{$3.7813 \pm 0.0071$}
\def\bbcg{$0.0466 \pm 0.0010$}
\def\ddca{$0.1451 \pm 0.0005$}
\def\ddcb{$3.0891 \pm 0.0062$}
\def\ddcg{$0.0464 \pm 0.0010$}
\def\eeaa{$0.1461 \pm 0.0006$}
\def\eeab{$2.7565 \pm 0.0070$}
\def\eeag{$0.0467 \pm 0.0009$}
\def\ddaa{$0.1443 \pm 0.0005$}
\def\ddab{$3.1160 \pm 0.0062$}
\def\ddag{$0.0474 \pm 0.0010$}
\def\ddqa{$0.1442 \pm 0.0005$}
\def\ddqb{$3.1154 \pm 0.0062$}
\def\ddqg{$0.0475 \pm 0.0010$}
\def\ddfa{$0.1440 \pm 0.0005$}
\def\ddfb{$3.0901 \pm 0.0061$}
\def\ddfg{$0.0313 \pm 0.0010$}
\def\eeqa{$0.1458 \pm 0.0006$}
\def\eeqb{$2.7526 \pm 0.0075$}
\def\eeqg{$0.0533 \pm 0.0009$}
\def\eena{$0.1417 \pm 0.0004$}
\def\eenb{$3.0770 \pm 0.0050$}
\def\eeng{$0.0006 \pm 0.0007$}

\def\ddfw{$-1.0201 \pm 0.0046$}
\def\bbqw{$-0.9896 \pm 0.0048$}
\def\eeaw{$-0.9572 \pm 0.0045$}
\def\ddaw{$-0.9777 \pm 0.0047$}
\def\ddqw{$-0.9828 \pm 0.0047$}
\def\bbcw{$-1.0105 \pm 0.0046$}
\def\bbaw{$-0.9961 \pm 0.0048$}
\def\ddcw{$-1.0086 \pm 0.0046$}
\def\eeqw{$-0.9545 \pm 0.0045$}
\def\eenw{$-0.9943 \pm 0.0040$}
\def\bbfw{$-1.0237 \pm 0.0046$}

\def\FOUNDxkweightone{$2.943 \pm 1.3$}
\def\FOUNDxkmeanone{$-0.972 \pm 0.506$}
\def\FOUNDxkstdone{$0.97 \pm 0.554$}

\def\FOUNDxlweightone{$3.36 \pm 1.158$}
\def\FOUNDxlmeanone{$-1.167 \pm 0.153$}
\def\FOUNDxlstdone{$0.493 \pm 0.192$}

\def\FOUNDxmweightone{$3.317 \pm 1.179$}
\def\FOUNDxmmeanone{$-1.152 \pm 0.097$}
\def\FOUNDxmstdone{$0.471 \pm 0.101$}

\def\FOUNDxnweightone{$3.342 \pm 1.162$}
\def\FOUNDxnmeanone{$-1.173 \pm 0.092$}
\def\FOUNDxnstdone{$0.559 \pm 0.103$}

\def\FOUNDxoweightone{$3.32 \pm 1.179$}
\def\FOUNDxomeanone{$-1.181 \pm 0.097$}
\def\FOUNDxostdone{$0.522 \pm 0.1$}

\def\FOUNDxpweightone{$3.257 \pm 1.215$}
\def\FOUNDxpmeanone{$-1.264 \pm 0.152$}
\def\FOUNDxpstdone{$0.618 \pm 0.256$}

\def\LOWZxjweightone{$1.874 \pm 1.023$}
\def\LOWZxjmeanone{$-1.485 \pm 0.644$}
\def\LOWZxjstdone{$1.438 \pm 0.454$}

\def\LOWZxkweightone{$2.402 \pm 1.125$}
\def\LOWZxkmeanone{$-1.917 \pm 0.342$}
\def\LOWZxkstdone{$0.835 \pm 0.52$}

\def\LOWZxlweightone{$3.271 \pm 1.181$}
\def\LOWZxlmeanone{$-1.884 \pm 0.102$}
\def\LOWZxlstdone{$0.502 \pm 0.107$}

\def\LOWZxmweightone{$3.371 \pm 1.172$}
\def\LOWZxmmeanone{$-1.836 \pm 0.163$}
\def\LOWZxmstdone{$0.458 \pm 0.169$}

\def\LOWZxnweightone{$3.374 \pm 1.179$}
\def\LOWZxnmeanone{$-1.816 \pm 0.058$}
\def\LOWZxnstdone{$0.432 \pm 0.043$}

\def\LOWZxoweightone{$3.365 \pm 1.186$}
\def\LOWZxomeanone{$-1.78 \pm 0.055$}
\def\LOWZxostdone{$0.433 \pm 0.041$}

\def\LOWZxpweightone{$3.333 \pm 1.161$}
\def\LOWZxpmeanone{$-1.788 \pm 0.059$}
\def\LOWZxpstdone{$0.452 \pm 0.046$}

\def\LOWZxqweightone{$3.362 \pm 1.165$}
\def\LOWZxqmeanone{$-1.678 \pm 0.064$}
\def\LOWZxqstdone{$0.451 \pm 0.06$}

\def\LOWZxrweightone{$3.326 \pm 1.181$}
\def\LOWZxrmeanone{$-1.613 \pm 0.085$}
\def\LOWZxrstdone{$0.408 \pm 0.111$}

\def\LOWZxsweightone{$3.42 \pm 1.143$}
\def\LOWZxsmeanone{$-1.719 \pm 0.377$}
\def\LOWZxsstdone{$1.341 \pm 0.624$}

\def\LOWZxCjweighttwo{$3.239 \pm 1.204$}
\def\LOWZxCjmeantwo{$0.402 \pm 0.181$}
\def\LOWZxCjstdtwo{$0.595 \pm 0.153$}

\def\LOWZxCkweighttwo{$3.254 \pm 1.169$}
\def\LOWZxCkmeantwo{$0.273 \pm 0.142$}
\def\LOWZxCkstdtwo{$0.693 \pm 0.123$}

\def\LOWZxClweighttwo{$2.265 \pm 1.144$}
\def\LOWZxClmeantwo{$0.418 \pm 0.635$}
\def\LOWZxClstdtwo{$0.805 \pm 0.282$}

\def\LOWZxCmweighttwo{$1.46 \pm 0.579$}
\def\LOWZxCmmeantwo{$0.24 \pm 0.116$}
\def\LOWZxCmstdtwo{$0.816 \pm 0.118$}

\def\LOWZxCnweighttwo{$1.367 \pm 0.665$}
\def\LOWZxCnmeantwo{$0.495 \pm 0.603$}
\def\LOWZxCnstdtwo{$0.866 \pm 0.333$}

\def\LOWZxCoweighttwo{$1.366 \pm 0.539$}
\def\LOWZxComeantwo{$0.394 \pm 0.108$}
\def\LOWZxCostdtwo{$0.73 \pm 0.097$}

\def\LOWZxCpweighttwo{$1.471 \pm 0.592$}
\def\LOWZxCpmeantwo{$0.418 \pm 0.101$}
\def\LOWZxCpstdtwo{$0.691 \pm 0.111$}

\def\LOWZxCqweighttwo{$1.04 \pm 0.544$}
\def\LOWZxCqmeantwo{$0.536 \pm 0.127$}
\def\LOWZxCqstdtwo{$0.578 \pm 0.12$}

\def\LOWZxCrweighttwo{$0.917 \pm 0.481$}
\def\LOWZxCrmeantwo{$0.577 \pm 0.163$}
\def\LOWZxCrstdtwo{$0.729 \pm 0.202$}

\def\LOWZxCsweighttwo{$0.696 \pm 0.61$}
\def\LOWZxCsmeantwo{$1.023 \pm 0.827$}
\def\LOWZxCsstdtwo{$1.662 \pm 0.688$}

\def\FOUNDxCkweighttwo{$2.217 \pm 1.367$}
\def\FOUNDxCkmeantwo{$0.645 \pm 0.543$}
\def\FOUNDxCkstdtwo{$0.955 \pm 0.487$}

\def\FOUNDxClweighttwo{$1.819 \pm 0.99$}
\def\FOUNDxClmeantwo{$0.682 \pm 0.28$}
\def\FOUNDxClstdtwo{$0.654 \pm 0.334$}

\def\FOUNDxCmweighttwo{$1.285 \pm 0.743$}
\def\FOUNDxCmmeantwo{$0.731 \pm 0.278$}
\def\FOUNDxCmstdtwo{$0.631 \pm 0.319$}

\def\FOUNDxCnweighttwo{$1.694 \pm 1.011$}
\def\FOUNDxCnmeantwo{$0.942 \pm 0.189$}
\def\FOUNDxCnstdtwo{$0.423 \pm 0.308$}

\def\FOUNDxCoweighttwo{$1.048 \pm 0.732$}
\def\FOUNDxComeantwo{$0.896 \pm 0.276$}
\def\FOUNDxCostdtwo{$0.574 \pm 0.399$}

\def\FOUNDxCpweighttwo{$0.904 \pm 0.644$}
\def\FOUNDxCpmeantwo{$0.845 \pm 0.451$}
\def\FOUNDxCpstdtwo{$0.914 \pm 0.581$}

\def\LOWZcjmean{$-0.043 \pm 0.042$}
\def\LOWZcjstdl{$0.046 \pm 0.037$}
\def\LOWZcjstdr{$0.164 \pm 0.039$}

\def\LOWZckmean{$-0.072 \pm 0.02$}
\def\LOWZckstdl{$0.018 \pm 0.016$}
\def\LOWZckstdr{$0.151 \pm 0.035$}

\def\LOWZclmean{$-0.075 \pm 0.017$}
\def\LOWZclstdl{$0.014 \pm 0.012$}
\def\LOWZclstdr{$0.134 \pm 0.022$}

\def\LOWZcmmean{$-0.074 \pm 0.015$}
\def\LOWZcmstdl{$0.013 \pm 0.01$}
\def\LOWZcmstdr{$0.154 \pm 0.018$}

\def\LOWZcnmean{$-0.073 \pm 0.014$}
\def\LOWZcnstdl{$0.012 \pm 0.009$}
\def\LOWZcnstdr{$0.136 \pm 0.015$}

\def\LOWZcomean{$-0.072 \pm 0.02$}
\def\LOWZcostdl{$0.017 \pm 0.013$}
\def\LOWZcostdr{$0.132 \pm 0.017$}

\def\LOWZcpmean{$-0.067 \pm 0.023$}
\def\LOWZcpstdl{$0.021 \pm 0.015$}
\def\LOWZcpstdr{$0.138 \pm 0.022$}

\def\LOWZcqmean{$-0.044 \pm 0.034$}
\def\LOWZcqstdl{$0.036 \pm 0.022$}
\def\LOWZcqstdr{$0.12 \pm 0.029$}

\def\LOWZcrmean{$-0.019 \pm 0.039$}
\def\LOWZcrstdl{$0.043 \pm 0.027$}
\def\LOWZcrstdr{$0.094 \pm 0.037$}

\def\LOWZcsmean{$0.029 \pm 0.035$}
\def\LOWZcsstdl{$0.064 \pm 0.03$}
\def\LOWZcsstdr{$0.036 \pm 0.029$}

\def\FOUNDckmean{$0.01 \pm 0.049$}
\def\FOUNDckstdl{$0.071 \pm 0.03$}
\def\FOUNDckstdr{$0.218 \pm 0.066$}

\def\FOUNDclmean{$0.004 \pm 0.041$}
\def\FOUNDclstdl{$0.06 \pm 0.031$}
\def\FOUNDclstdr{$0.192 \pm 0.082$}

\def\FOUNDcmmean{$-0.004 \pm 0.04$}
\def\FOUNDcmstdl{$0.064 \pm 0.025$}
\def\FOUNDcmstdr{$0.2 \pm 0.055$}

\def\FOUNDcnmean{$0.01 \pm 0.047$}
\def\FOUNDcnstdl{$0.067 \pm 0.027$}
\def\FOUNDcnstdr{$0.181 \pm 0.057$}

\def\FOUNDcomean{$0.011 \pm 0.051$}
\def\FOUNDcostdl{$0.069 \pm 0.028$}
\def\FOUNDcostdr{$0.168 \pm 0.062$}

\def\FOUNDcpmean{$0.024 \pm 0.05$}
\def\FOUNDcpstdl{$0.072 \pm 0.029$}
\def\FOUNDcpstdr{$0.085 \pm 0.043$}

\def\LOWZcCjmean{$-0.042 \pm 0.044$}
\def\LOWZcCjstdl{$0.043 \pm 0.038$}
\def\LOWZcCjstdr{$0.163 \pm 0.044$}

\def\LOWZcCkmean{$-0.069 \pm 0.023$}
\def\LOWZcCkstdl{$0.02 \pm 0.019$}
\def\LOWZcCkstdr{$0.157 \pm 0.041$}

\def\LOWZcClmean{$-0.067 \pm 0.018$}
\def\LOWZcClstdl{$0.015 \pm 0.013$}
\def\LOWZcClstdr{$0.128 \pm 0.025$}

\def\LOWZcCmmean{$-0.065 \pm 0.016$}
\def\LOWZcCmstdl{$0.013 \pm 0.011$}
\def\LOWZcCmstdr{$0.147 \pm 0.018$}

\def\LOWZcCnmean{$-0.062 \pm 0.015$}
\def\LOWZcCnstdl{$0.012 \pm 0.01$}
\def\LOWZcCnstdr{$0.126 \pm 0.016$}

\def\LOWZcComean{$-0.061 \pm 0.018$}
\def\LOWZcCostdl{$0.016 \pm 0.012$}
\def\LOWZcCostdr{$0.118 \pm 0.017$}

\def\LOWZcCpmean{$-0.061 \pm 0.021$}
\def\LOWZcCpstdl{$0.017 \pm 0.014$}
\def\LOWZcCpstdr{$0.126 \pm 0.022$}

\def\LOWZcCqmean{$-0.044 \pm 0.031$}
\def\LOWZcCqstdl{$0.028 \pm 0.021$}
\def\LOWZcCqstdr{$0.111 \pm 0.028$}

\def\LOWZcCrmean{$-0.024 \pm 0.039$}
\def\LOWZcCrstdl{$0.033 \pm 0.025$}
\def\LOWZcCrstdr{$0.09 \pm 0.037$}

\def\LOWZcCsmean{$0.019 \pm 0.033$}
\def\LOWZcCsstdl{$0.047 \pm 0.03$}
\def\LOWZcCsstdr{$0.035 \pm 0.031$}

\def\FOUNDcCkmean{$-0.046 \pm 0.05$}
\def\FOUNDcCkstdl{$0.048 \pm 0.034$}
\def\FOUNDcCkstdr{$0.212 \pm 0.06$}

\def\FOUNDcClmean{$-0.069 \pm 0.041$}
\def\FOUNDcClstdl{$0.037 \pm 0.029$}
\def\FOUNDcClstdr{$0.243 \pm 0.041$}

\def\FOUNDcCmmean{$-0.07 \pm 0.034$}
\def\FOUNDcCmstdl{$0.033 \pm 0.023$}
\def\FOUNDcCmstdr{$0.211 \pm 0.054$}

\def\FOUNDcCnmean{$-0.055 \pm 0.039$}
\def\FOUNDcCnstdl{$0.042 \pm 0.026$}
\def\FOUNDcCnstdr{$0.206 \pm 0.057$}

\def\FOUNDcComean{$-0.028 \pm 0.048$}
\def\FOUNDcCostdl{$0.061 \pm 0.031$}
\def\FOUNDcCostdr{$0.198 \pm 0.067$}

\def\FOUNDcCpmean{$0.026 \pm 0.049$}
\def\FOUNDcCpstdl{$0.086 \pm 0.031$}
\def\FOUNDcCpstdr{$0.115 \pm 0.08$}

\usepackage{amsmath}
\usepackage[hang]{footmisc}

\usepackage{floatrow}
\usepackage[title]{appendix}

\begin{document}

\preprint{APS/123-QED}

\title{Improved Treatment of Host-Galaxy Correlations in Cosmological Analyses With Type Ia Supernovae}

\submitted{}

\author{Brodie Popovic\footnotemark[1], Dillon Brout\footnotemark[2,3], Richard Kessler\footnotemark[4,5], Dan Scolnic\footnotemark[1], Lisa Lu \footnotemark[6]}

\affiliation{$^1$Department of Physics, Duke University, Durham, NC, 27708, USA.}
\affiliation{$^2$ Center for Astrophysics, Harvard \& Smithsonian, 60 Garden Street, Cambridge, MA 02138, USA}
\affiliation{$^3$ NASA Einstein Fellow}
\affiliation{$^4$Department of Astronomy and Astrophysics, The University of Chicago, Chicago, IL 60637, USA.} \affiliation{$^5$Kavli Institute for Cosmological Physics, University of Chicago, Chicago, IL 60637, USA.}
\affiliation{$^6$Independent Scholar}

\date{\today}

\begin{abstract}
Improving the use of Type Ia supernovae (SNIa) as standard candles requires a better approach to incorporate the relationship between SNIa and the properties of their host galaxies. Using a spectroscopically-confirmed sample of $\sim$1600 SNIa, we develop the first empirical model of underlying populations for SNIa light-curve properties that includes their dependence on host-galaxy stellar mass. These populations are important inputs to simulations that are used to model selection effects and correct distance biases within the BEAMS with Bias Correction (BBC) framework. Here we improve BBC to also account for SNIa-host correlations, and we validate this technique on simulated data samples. We recover the input relationship between SNIa luminosity and host-galaxy stellar mass (the mass step, $\gamma$) to within \gammabias mags, which is a factor of 5 improvement over the previous method that results in a $\gamma$-bias of ${\sim}0.02$. We adapt BBC for a novel dust-based model of intrinsic brightness variations, which results in a greatly reduced mass step for data ($\gamma = 0.017 \pm 0.008$), and for simulations ($\gamma =0.006 \pm 0.007$). Analysing simulated SNIa, the biases on the dark energy equation-of-state, $w$, vary from $\Delta w = 0.006(5)$ to $0.010(5)$ with our new BBC method; these biases are significantly smaller than the $0.02(5)$ $w$-bias 
using previous BBC methods that ignore SNIa-host correlations.
\end{abstract}

\newcommand{\siglow}{$\sigma_-$}
\newcommand{\sighigh}{$\sigma_+$}
\newcommand{\massbin}{$0.2 \times 10^{10} ~ M_{\odot}$}
\newcommand{\masswindow}{$ 1.2 \times 10^{10} ~ M_{\odot}$}
\newcommand{\size}{50 SNIa~}
\newcommand{\SNLSsize}{75 SNIa~}
\newcommand{\xtreme}{BBC-2020~}
\newcommand{\numsamples}{100}

\newcommand{\BiasCorStep}{\theta} 
\newcommand{\DmM}{\delta \mu_{\textrm{host}}}
\newcommand{\dmuBiasCor}{\delta\mu_{\textrm{bias}}}
\newcommand{\finterp}{f}
\newcommand{\mass}{$M_{\textrm{stellar}}$}
\newcommand{\zindex}{z_i}
\newcommand{\delbias}{\delta \mu_{\textrm{bias}}}
\newcommand{\delhost}{\delta \mu_{\textrm{host}}}
\newcommand{\wbias}{$w$-bias}

\maketitle

\section{\label{sec:Intro}Introduction}

\footnotetext[1]{Email: brodie.popovic@duke.edu}

{The standardisation of measurements of Type Ia Supernovae (SNIa) led to the discovery that the universe is expanding at an increasing rate \citep{Riess98, Perlmutter99}.} A possible cause of this expansion, called `dark energy', remains an unsolved mystery in cosmology to this day. In the intervening decades since the discovery of dark energy, SNIa cosmology has grown from sample sizes of tens to thousands. When combined with constraints from the Cosmic Microwave Background, the statistical uncertainties in measurements of the dark energy equation-of-state parameter, $w$, are on the order of $\sim 0.04$, \citep{Scolnic18, Jones18, Brout19b}. Therefore, understanding systematic uncertainties on the level of $\sim 0.01$ is needed to better measure the nature of dark energy. To further reduce systematic uncertainties, here we improve the treatment of the correlation between supernova properties and host-galaxy properties. 

Most SNIa cosmology analyses use the SALT2 \citep{Guy10} framework that relies on two parameters to standardise the SNIa brightness: a colour ($c$) describing the wavelength-dependent luminosity and a light-curve stretch ($x_1$) describing the luminosity dependence on light-curve duration. In addition to these SALT2 luminosity correlations, studies have shown a correlation between the standardised brightness of SNIa and properties of their host-galaxy, such as mass (e.g. \citealp{Sullivan10}), Star Formation Rate (e.g. \citealp{Uddin17}), as well as other properties \citep{Rose2020}. While $c$ and $x_1$ are rigorously included in the SALT2 model, the correlation with host properties is typically included as an ad-hoc correction to the supernova brightness. Additionally, correlations between $c$ and host-properties as well as $x_1$ and host-properties have been ignored when this ad-hoc correction is applied. In this analysis, we focus on host-galaxy stellar mass (\mass) as a representative host-galaxy property. 

To predict and correct for biases in the measurement of distance modulus values from standardisation methods, analyses have increasingly relied on simulations \citep{Kessler09, Betoule14, Scolnic18, Brout19b}. These simulations used the publicly available SuperNova ANAlysis package (SNANA: \citealp{SNANA}); a detailed description of the simulations is given in \cite{Kessler19}. To simulate SNIa samples, simulations rely on measurements of underlying populations of $c$ and $x_1$. Early SNIa cosmology analyses tuned these populations based on visual comparisons of fitted $c$ and $x_1$ distributions between data and simulations and lacked rigorous procedure. This motivated \citealp{Scolnic16} (SK16) to formalise a method for determining the underlying populations. By utilizing large simulations, SK16 determined the underlying populations by tracking how selection effects, noise and intrinsic scatter cause populations to `migrate' from underlying distributions to observed distributions. Here, we improve on SK16 by determining underlying $c$ and $x_1$ populations as a function of \mass. {Our population fitting code is publicly available\footnote{https://github.com/bap37/ParentPops}, along with sample inputs.\footnote{T.B.D}.}

Simulated samples can be used to determine redshift-dependent bias corrections, as done in the analysis for SDSS-II \citep{Kessler09} and for the Joint Lightcurve Analysis (JLA; \citealp{Betoule14}). Subtle $c$ and $x_1$ biases were shown in SK16 for simulated distance moduli, suggesting that 1-dimensional redshift-dependent bias corrections are not sufficient. This motivated \cite{Kessler16} to develop the BEAMS with Bias Corrections (BBC) method, which corrects for observed light-curve fit parameters based on rigorous simulations with corrections computed in 5-dimensional parameter space (redshift, colour, stretch, colour-luminosity relationship, and stretch-luminosity relationship). Hereafter, we refer to this method as BBC5D, which has been used in Dark Energy Survey 3 year analysis (DES3YR; \citealp{DES3YR, Brout19b}), analyses of the Pan-STARRs sample \citep{Scolnic18, Jones18}, and in a re-analysis of the Sloan Digital Sky Survey-II (SDSS) \citep{Popovic19}. We refer to the redshift-only correction as BBC1D. A summary of these methods is presented in Table \ref{tab:Ref}. 

Using SNANA to simulate DES3YR with a known input value of $\gamma$, \cite{Smith20} showed that the BBC5D-recovery of $\gamma$ from an input value is significantly biased when correlations between $c$/$x_1$ and \mass~ are included in the simulation. To address the $\gamma$ bias found in \cite{Smith20}, we improve the BBC formalism from BBC5D to BBC7D by introducing two new dimensions in the bias corrections.

\begin{table*}[t]
\scalebox{1}{%
\begin{tabular}{c|c|c|c}
    Method & Bias-Correction & Comment & Citation \\
     & Dependence & & \\
    \hline
    BBC1D & $z$ & Biascor grid of $z$: correct $\mu (z).$ & \cite{Marriner11} \\
     &  &  & \cite{Kessler16}  \\
    \hline
    BBC5D & $\{z,x_1,c,\alpha, \beta \}$ &  Biascor grid of $\alpha, \beta, c, x_1, z$: correct $m_B(z), c$ and $x_1$ & \cite{Kessler16} \\
    \hline
    BBC7D & $\{z,x_1,c,\alpha, \beta, \theta, M_{\textrm{stellar}} \}$ & Biascor grid of $\alpha, \beta, \c, x_1, z, \gamma, M_{\rm stellar}$: correct $m_B (z), c$ and $x_1$. & This Work\\
    \hline
    BBC-BS20 & $\{z,x_1,c, M_{\textrm{stellar}} \}$ & Biascor based on BS20 model, correct $\mu(z)$ in grid of $c, x_1, z ,M_{\rm stellar}$ & This Work \\
    \hline
\end{tabular}%
}
\caption{Breakdown of different bias correcting methods.}
\label{tab:Ref}
\end{table*}

\citealp{BS20}, hereafter BS20, present a new explanation for the mass step $\gamma$ as a consequence of varying extinction ratios for high and low mass host galaxies. BS20 extended the SALT2 formalism to include two sources of colour variation: intrinsic and dust. This formalism, however, is incompatible with BBC5D and BBC7D; therefore we adapt BBC to work with BS20 (BBC-BS20).

Here, we address the concerns about corrections for the mass step raised by \cite{Smith20}, and confirm the origin of the mass step presented in BS20. Section \ref{sec:Data} provides an overview of the data used in this work. Section~\ref{sec:Sims&Analysis} is an overview of population-fitting techniques and selection effects. In Section~\ref{sec:Modeling} we review the methodology introduced in previous studies and expand on them to track evolution with host-galaxy properties. Section~\ref{sec:BBC} details changes to the BBC formalism, the efficacy of which is presented in Section~\ref{sec:Results}. Finally, discussion and conclusions are presented in Section~\ref{sec:Conclusion}.

\section{Data}\label{sec:Data}

In this analysis, we compile light-curves from 6 SNIa samples that have been spectroscopically confirmed. The samples used here are from the Sloan Digital Sky Survey-II (SDSS; \citealp{York2000, Sako18}), the Supernova Legacy Survey (SNLS; \citealp{Sullivan10}), Pan-STARRs (PS1; \citealp{Jones18, Scolnic18}), the Foundation Supernova Survey \citep{Foley17} and the Dark Energy Survey (DES; \citealp{Brout19a}). The low-redshift (Low-z) supernovae include samples from the Carnegie Supernova Project (CSP) \citep{Stritzinger11} and Harvard-Smithsonian Center for Astrophysics (CfA3-4) (\citealp{Hicken09b, Hicken09a, Hicken12}). 

\begin{figure}
\includegraphics[width=8cm]{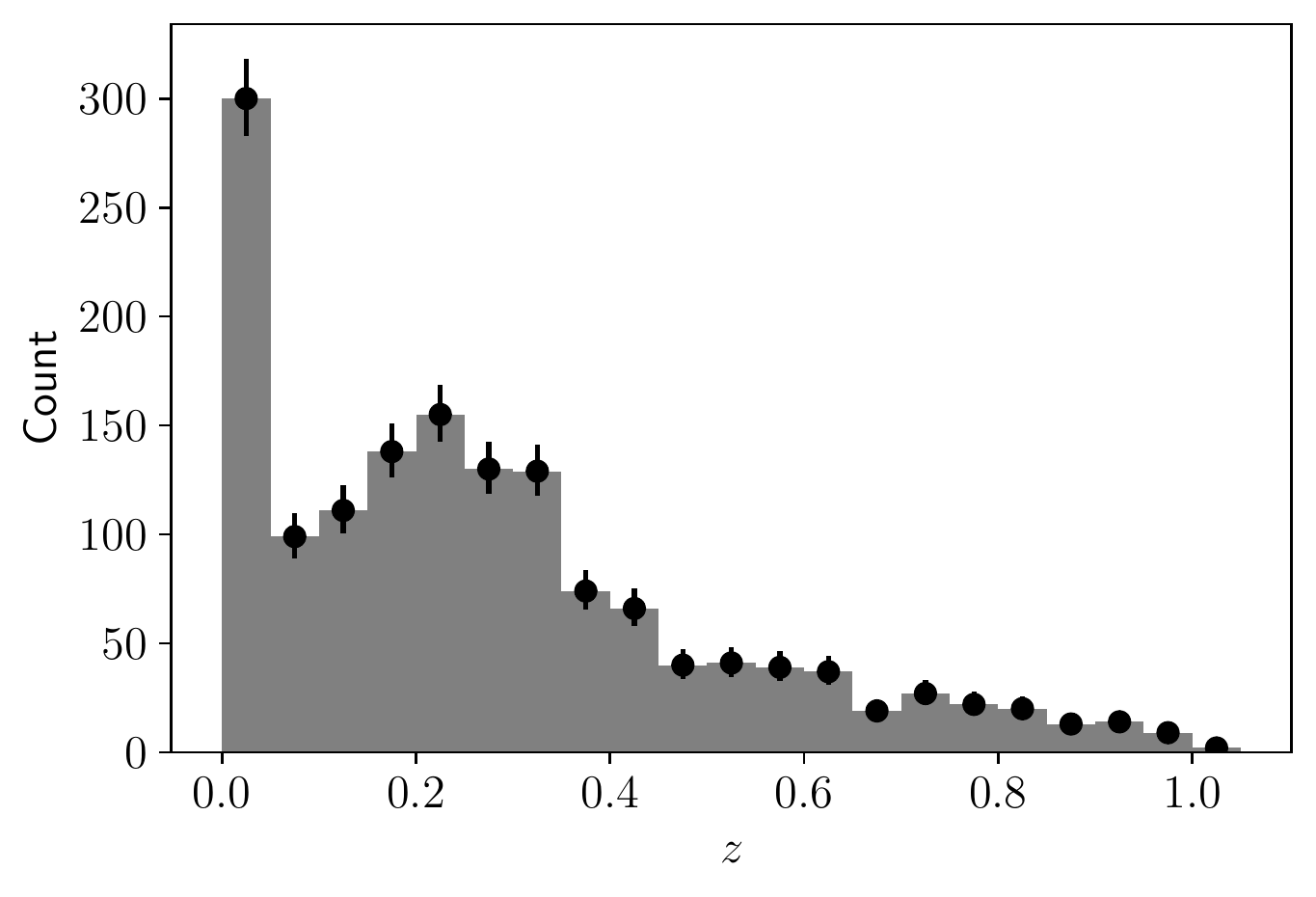}
\caption{The redshift distribution of the combined Foundation, Low-z, SDSS, DES, SNLS, and PS1 data sample.}
\label{fig:z-dist}
\end{figure}

These samples are flux-calibrated to the SuperCal system \citep{Scolnic15}. Host-galaxy masses are taken from past analyses, notably we use the masses for SDSS, PS1, and SNLS provided in the Pantheon sample \citep{Scolnic18}. For Foundation, we use the masses provided by \cite{Jones18}. For DES, we use updated masses provided by \cite{Smith20} and \cite{Wiseman20}. These analyses derive host-galaxy masses from SED fitting to broadband photometry. 
To measure parent populations, we apply the following selection requirements (cuts):

\begin{itemize}
    \item light-curve fit probability $P_{\textrm{fit}} > 0.01$ that varies for each survey
    \item fitted SNIa colour $|c| < .3$
    \item fitted SNIa colour uncertainty $\sigma_c < .2$
    \item fitted SNIa stretch $|x_1| < 3$
    \item fitted SNIa stretch uncertainty $\sigma_{x1} < 1$
    \item fitted SNIa time-of-peak-brightness error $\sigma_{t0} < 2$
    \item at least five observations
    \item at least one observation before peak brightness
    \item at least one observation after peak brightness
\end{itemize}

However, for Low-z, we drop the $c$/$x_1$ uncertainty cuts to increase statistics. A summary of the total SNIa for each sample is given in Table \ref{tab:cuts}. The resulting redshift distribution is shown in Figure \ref{fig:z-dist}. The $c$, $x_1$, and \mass~ distributions for our combined data sample are presented in Figure \ref{fig:obs-dist}, along with the mean values for $c$ and $x_1$ as a function of \mass. 

\begin{table}[h]
    \centering
    \begin{tabular}{c|c}
         Sample &  \# SNIa  \\
         \hline
        Low-z & 270  \\ 
        \hline
        Found & 120  \\ 
        \hline
        SDSS &  385  \\ 
        \hline
        SNLS &  237  \\ 
        \hline
        DES &   227  \\ 
        \hline
        PS1 &   308  \\ 
        \hline
        Total & 1547  
    \end{tabular}
    \caption{Summary of Data Statistics}
    \label{tab:cuts}
\end{table}

\section{ Simulations and Analysis}\label{sec:Sims&Analysis}

\subsection{Simulations}\label{sec:Sims&Analysis:subsec:Sims}

Simulations of supernovae are needed to correct for biases arising from inefficiencies and from the light-curve fitting process. Bias corrections are needed to measure populations for stretch and colour, and to measure distances. Therefore we generate simulations to assess the impact of systematics on cosmological measurements. To this end, we use SNANA to simulate realistic samples of SNIa with appropriate noise, observing conditions, cadence, and detection efficiencies. We generate these simulations with the Flat $\Lambda$-CDM cosmology from \cite{Planck16}: $\Omega_{\textrm{M}} = 0.315$ (matter density at $z=0$) and $w = -1$.

\begin{figure}
\includegraphics[width=9cm]{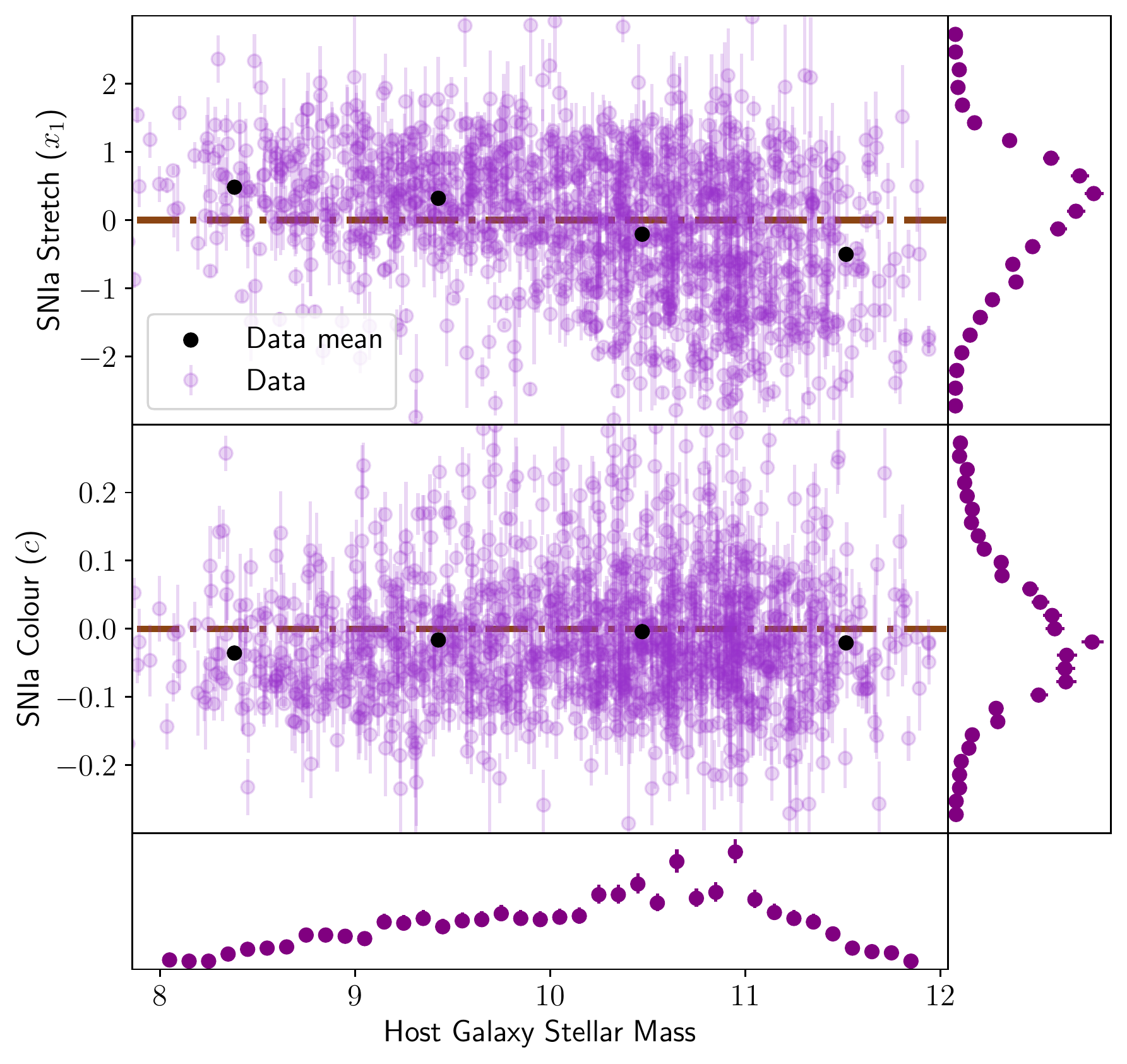}
\caption{The distributions of $c$ and $x_1$ for the combined Low-z, DES, PS1, Foundation, SNLS, and SDSS spectroscopic samples as a function of host-galaxy mass. Circles are the data; purple represent individual data points and black are the average value for that bin. The dash-dotted line is 0.  }
\label{fig:obs-dist}
\end{figure}

The SNANA simulation is presented in \cite{Kessler19}, and here we provide a brief overview. The simulation begins by generating a Spectral Energy Distribution and applying dimming from cosmic expansion, lensing, galactic extinction, and redshift effects. Next, telescope-specific information (PSFs, zero point, sky noise, cadence) is used to simulate measured fluxes and uncertainties in each passband. Finally, selection efficiencies from the detection pipeline and spectroscopic targeting are applied to select events for analysis. We modify the spectroscopic efficiency function, Eff$_{\textrm{spec}}$, for the SNLS survey to better fit the data (Appendix A1).

These simulations rely on previously derived frameworks that describe brightness variations among the SNIa population, which we call intrinsic scatter. Here, we use the G10 \citep{Guy10} and C11 \citep{Chotard11} scatter models, translated into SED variants in \cite{Kessler13}, along with the BS20 model.
G10 and C11 are spectral variation models that differ in the amount of variation ascribed to chromatic versus achromatic scatter.  For G10, roughly 70\% of scatter is achromatic and 30\% is chromatic, whereas for C11, roughly 25\% of scatter is achromatic and 75\% is chromatic. The BS20 model does not have an explicit SED variation model; instead it includes dust parameter populations that are not part of the SALT2 framework.

To accurately characterise our systematic and statistical uncertainties, we simulate \samples data-sized samples.

\subsection{Analysis}\label{sec:Sims&Analysis:subsec:Analysis}
We fit the light-curves using SNANA with the SALT2 model as developed in \cite{Guy10} and updated in \cite{Betoule14}. 
For each SNIa, the SALT2 fit gives four fitted parameters: $m_B$, the log of the fitted light-curve amplitude $x_0$; $x_1$, the stretch parameter corresponding to light-curve width,; $c$, the light-curve colour; and $t_0$, the time of peak brightness. 
From the fitted SALT2 parameters, we infer distance modulus values with a modified version of the Tripp distance estimator \citep{Tripp}. 
Following the BBC formalism in \cite{Kessler16}, the distance modulus ($\mu)$ is defined as:
\begin{equation}
    \mu=m_B + \alpha x_1 - \beta c - M_{\zindex} +\delhost + \delbias \label{eq:Tripp}
\end{equation}
where $\alpha$ and $\beta$ are global nuisance parameters relating to stretch and colour respectively. $M_{\zindex}$ is the distance offset in discrete redshift bins denoted by $\zindex$. $\delhost$ is the luminosity correction for the mass step, and is defined as:
\begin{equation}
\delhost = \gamma \times (1 + e^{(X_{\textrm{stellar}} - S)/ \tau_X})^{-1} - \frac{\gamma}{2}~, \label{eq:BCmu2}
\end{equation}
where $\gamma$ is the magnitude of the SNIa luminosity difference between SNIa in high and low mass galaxies, $X_{\textrm{stellar}}$ is $\log_{10}$\mass, $S \sim 10$ is the step location, and $\tau_X$ is the width of the step. Finally, $\delbias$ is the distance bias correction. As summarised in Table 1, for BBC1D, $\delbias$ is determined in bins of redshift using simulations with a fixed $\alpha$ and $\beta$. For higher-dimensional corrections (BBC5D/BBC7D/BBC-BS20), the corrections are applied in bins of redshift along with other SNIa and host galaxy properties.

Furthermore, the total distance modulus error $\sigma^2$ is described as 

\begin{equation}\label{eq:TrippErr}
    \sigma^2 = \sigma^2_N + \sigma^2_{\textrm{lens}} + \sigma^2_{\sigma z} +  \sigma^2_{\textrm{int}} 
\end{equation}
where $\sigma_N$ includes the uncertainties from SALT2 parameters and their covariances (see Equation 3 in \cite{Kessler16}), $\sigma_{\textrm{lens}} = 0.055z$ is the uncertainty from weak gravitational lensing \citep{Scolnic18}, $\sigma_{\sigma z}$  is the uncertainty from peculiar velocity \citep{Carrick15}, and $\sigma_{\textrm{int}}$ is the intrinsic scatter contribution determined in the BBC fitting process.

The BBC fit maximises a likelihood that includes separate terms for SNIa and core collapse (CC) SNe. Because our sample includes only spectroscopically confirmed events, the CC term is excluded. The BBC fit determines $\alpha$, $\beta$, $\gamma$, $\sigma_{\textrm{int}}$ over the entire redshift range, and a distance modulus in each redshift bin ($\zindex$). Following \cite*{bhs20}, because we are not assessing the covariances of systematics, a $z$-binned Hubble diagram results in equivalent cosmological constraints as that of an unbinned set.  These binned distances are combined with priors in cosmological fits. Here, $\Omega_{\textrm{M}}$ and $w$ are obtained with ``{\tt wfit}'', a $\chi^2$ minimisation program using MINUIT \citep{MINUIT}. We use a Gaussian $\Omega_{\textrm{M}}$ prior with a mean of 0.315 and width of $0.01$.

\section{Modeling SNIa Populations With Host Galaxy Correlations}\label{sec:Modeling}
Here, we determine the underlying correlations between SNIa parameters and host-galaxy properties given different models of intrinsic scatter (G10, C11: Section~\ref{sec:Data}; BS20: Section~\ref{sec:Modeling:subsec:BS20-Ext}). In Section~\ref{sec:Modeling:subsec:Review}, we review the methodology from SK16, who developed an underlying parent population method to ensure that resulting simulated distributions of $c$ and $x_1$ match the data. Naively, we would extend SK16 to a three dimensional population that includes \mass, however, the SK16 method relies on a robust model of simulated uncertainties that has been validated for $c$ and $x_1$, but not for \mass. Therefore, we use the SK16 method for $c$ and $x_1$, and introduce a different method for \mass~ which does not rely on modeling \mass~ uncertainties.

We fit our underlying parameter populations separately for Low-z, Foundation, and the combined high-$z$ collection of DES, SDSS, SNLS, and PS1. As with SK16, we find notable differences between low and high redshift surveys. The difference in \mass\ distributions and selection effects between Low-z and Foundation (Figure 3 in \citealp{Jones19}) motivate the separate fits.

We derive the dependence of underlying parameter populations on host-galaxy mass for the G10 and C11 models (\ref{sec:Modeling:subsec:Review} and \ref{sec:Modeling:subsec:Extend}). In Section~\ref{sec:Modeling:subsec:BS20} and \ref{sec:Modeling:subsec:BS20-Ext}, we review the BS20 framework and describe improvements to modeling the parent populations.

\subsection{Review of Method to Determine Uncorrelated Parent Populations}\label{sec:Modeling:subsec:Review}
Following SK16, we extend their stretch population description for high-redshift surveys (PS1, SDSS, DES, and SNLS) by replacing the asymmetric Gaussian with an asymmetric generalised normal distribution that depends on four parameters,

\begin{equation}\label{eq:AGauss}
P(x_1) = 
\begin{cases}
      e^{(-|x_1-\overline{x}_1|^n / n\sigma_-^n )} & \textrm{if} ~x_1 \leq \overline{x}_1 \\
      e^{|-|x_1-\overline{x}_1|^n / n\sigma_+^n )} & \textrm{if}~ x_1 > \overline{x}_1 
    \end{cases}
\end{equation}
where $\overline{x}_1$ is the value at peak probability of the asymmetric generalised normal distribution, $n$ is the shape, and \siglow ~ and \sighigh ~ are the width parameters for negative and positive $x_1$ values, respectively. The motivation for this distribution is discussed in Section \ref{sec:Results:subsec:ParentPops}. For $n = 2$, Equation \ref{eq:AGauss} reduces to the three parameter asymmetric Gaussian distribution used in SK16. This parametrisation is also used to describe the SNIa colour $c$. 

For Low-z and Foundation, the stretch distribution is double-peaked. We therefore take an alternative approach following \cite{Scolnic18} and use a double Gaussian model,
\begin{equation}
    P(x_1) = A_1 \times e^{(-|x_1-\overline{x_1}_1|^2 / 2\sigma_1^2 )} + A_2 \times e^{(-|x_1-\overline{x_1}_2|^2 / 2\sigma_2^2 )}
    \label{eq:DGauss}
\end{equation}
where $A_i$ is the weight, $\overline{x_1}_i$ is the mean value, and $\sigma_i$ is the standard deviation of the respective Gaussian. 

A simulation with a flat distribution of true colour and true stretch is generated, and SALT2 light-curve fits determine the measured colour and measured stretch for the SNIa that pass light-curve quality cuts. The true and measured values from these flat simulations are used to compute a migration matrix ($X$ for stretch and $C$ for colour) that captures the migration from the input distribution to the observed distribution through selection effects, measurement noise and intrinsic scatter. Each component of the matrix, $X_{ij}$, describes the likelihood that a true value in an input stretch bin $i$ $(x_{1_i})$ migrates to a measured value in stretch bin $j$ $(x_{1_j})$, and similarly for colour. 

For an underlying stretch population, $P(x_1)$, we define a binned distribution $\Vec{P}_{x}$ with components $P_{xi}$. SK16 defines $\Vec{\Delta}_{x}$, the data-simulation difference vector, as 

\begin{gather}\label{eq:Delta_c}
    \Vec{\Delta}_{x_1} = 
\begin{bmatrix}
o_{x1}  \\
o_{x2}  \\
\vdots    \\
o_{xn}  
\end{bmatrix}
- 
\begin{bmatrix}
X_{1,1} & X_{1,2} & \cdots & X_{1,n} \\
X_{2,1} & X_{2,2} & \cdots & X_{2,n} \\
\vdots  & \vdots  & \ddots & \vdots  \\
X_{d,1} & X_{d,2} & \cdots & X_{d,n} 
\end{bmatrix}
\times
\begin{bmatrix}
P_{x1}  \\
P_{x2}  \\
\vdots    \\
P_{xn}  
\end{bmatrix}
\end{gather}
where the measured distribution vector $\Vec{o}_x$ has the same binning as $X$ and $\Vec{P}_x$. 

For the $\chi^2$ calculation, the associated data error vector is $\Vec{e}_x = [e_{x1}, e_{x2}, ..., e_{xn}]$, where $e_{x_i} = \sqrt{o_{x_i}}$ for ${o_{x_i}} > 0$ and $e_{x_i} = 1$ for ${o_{x_i}} = 0$; while not technically correct for a Poisson distribution \citep{Baker}, SK16 has shown this this error approximation is sufficient. The four parameters that describe $P(x_1)$ are determined by minimising the $\chi_x^2$ defined as
\begin{equation}\label{eq:chi2}
    \chi_x^2 = \sum_{i=1}^{n} \left (\frac{\Delta_{xi}}{e_{xi}}\right )^2~.
\end{equation}
For the colour distribution, $\chi^2_c$  is defined similarly using $\Vec{\Delta}_{c}$ and $\Vec{e}_c$. 

SK16 performed a grid search for their parameters, where here we use a Monte Carlo minimisation procedure using the {\tt emcee} python package \citep{emcee}. We compare our population parameters to those of SK16 and find that we replicate their results to within $1\sigma$ for parameters describing the colour and stretch distributions
(Equation \ref{eq:AGauss} with $n=2$). We find that this assumption works well for the $c$ population, and therefore we fix $n=2$ (see Appendix A2). For $x_1$, however, we find that fixing $n=3$ works better and the resulting $\chi_{x1}^2$ is smaller by $\sim 3$ compared to $n=2$.

\subsection{Extending Parent Populations to Include Mass Dependence}\label{sec:Modeling:subsec:Extend}

The ideal approach for including mass-dependent correlations is to replace $\vec{o}_{x1}$ in Equation \ref{eq:Delta_c} with a 2-dimensional array of stretch and host-galaxy mass, $\vec{o}_{x1,M}$, where the subscript is $M = $ \mass. Similarly, the migration matrix $X$ and probability vector $\vec{P}_x$ would also be extended to include \mass.

However, the \mass\ measurements lack a well defined uncertainty and therefore the migration matrix is not as well determined in the \mass\ dimension as it is for $c$ or $x_1$. We therefore assume that the measured \mass~ is the true \mass. To reduce the dependence on \mass~ uncertainties, we implement a 2 step process. First, we fit the parent populations by minimising $\chi_x^2$ and $\chi_c^2$ (Eq. \ref{eq:chi2}) in \mass\ bins. Ideally we would perform this minimisation in small \mass\ bins, however, the statistics of SNIa per bin is insufficient. As a compromise, we fit in relatively large \mass\ bins of \masswindow\ but use a small step size of \massbin. Although the \mass\ bins are strongly correlated, we have used simulated data samples to validate our method. In the second step, we re-weight the simulated \mass\ distribution to match the data.

\subsection{REVIEW OF BS20}\label{sec:Modeling:subsec:BS20}

BS20 observed a significant ($>10 \sigma$) colour-dependent Hubble scatter, and a $5 \sigma$ colour-dependence on the mass step ($\gamma$). They model this effect with simulations using the SALT2 model combined with host galaxy dust. They define 3 contributions to the observed colour,
\begin{equation}\label{eq:c-component}
    c_{\textrm{obs}} = c_{\textrm{int}} + E_{\textrm{dust}} + \epsilon_{\rm noise} 
\end{equation}
where $c_{\textrm{int}}$ is the intrinsic SNIa colour, $E_{\textrm{dust}}$ is the dust extinction, and $\epsilon_{\rm noise}$ is measurement noise. These parameters contribute to a total change in true SNIa brightness of 
\begin{equation}
    \Delta m_{B, \textrm{obs}} = \beta_{\textrm{SN}} \times c_{\textrm{int}} + R_V \times E_{\textrm{dust}}
    \label{eq:BS20-deltamb}
\end{equation}
where $\beta_{\textrm{SN}}$ is introduced as the correlation coefficient between the intrinsic colour and SNIa luminosity, and $R_V$ is the dust extinction ratio. Components of the BS20 model that are intrinsic to the SNIa, i.e. $c_{\rm int}$ and $\beta_{\rm SN}$, are assumed to be independent of \mass. Thus, BS20 determine the following parameters describing the distributions of dust and colour: Gaussian distribution of intrinsic color ($\bar{c}_{\rm int}$, $\sigma_{c \textrm{int}}$), Gaussian distribution of intrinsic color-luminosity coefficient ($\bar{\beta}_{\rm SN}$ \& $\sigma_{\beta_{\rm SN}}$), and Gaussian distribution of dust extinction ratios ($\bar{R}_V$ \& $\sigma_{Rv}$) for low and high mass, and exponential distribution of the dust extinction ($\tau_E$) for low and high mass.

The BS20 parameters are determined in a forward-modeling fitting process by minimising the $\chi_{\rm TOT}^2$ in Eq. 8 of BS20. The $\chi_{\rm TOT}^2$ includes constraints enforcing consistency between data and simulations for: the colour distribution, the Hubble diagram scatter versus colour, the BBC-fitted Hubble residual versus colour, and BBC-fitted $\beta$. BS20 fixes the $\alpha$ in the simulations to be that obtained from the dataset using BBC1D, and fix $\gamma = 0$. 

\subsection{Upgrades to Parent Populations for Dust Based Scatter Models}\label{sec:Modeling:subsec:BS20-Ext}

For the BS20 model, we derive the $x_1$ populations following the method discussed in Section~\ref{sec:Modeling:subsec:BS20-Ext}. However, we keep the same colour distribution presented in BS20 because their dust model parameterisation is incompatible with the SK16 method. While this upgrade to the BS20 population parameters is minimal, there are significant upgrades to the BBC formalism presented in Section~\ref{sec:BBC:subsec:BS20}. These upgrades to the BBC formalism are essential because BS20 is incompatible with higher dimensional BBC procedures. A new and significantly improved formalism for determining the underlying population parameters for BS20 is under development and will be presented in a future work.

\section{Correcting Distance Biases}\label{sec:BBC}
Here we review the BBC methodology for correcting distance biases (Section~\ref{sec:BBC:subsec:Review}) and describe improvements. In Section~\ref{sec:BBC:subsec:Ext}, we detail the improvements to account for correlations between SNIa brightness and host-galaxy properties. Finally, in Section~\ref{sec:BBC:subsec:BS20}, we describe improvements to the BBC formalism required for bias corrections using the BS20 model.

\subsection{Review of 5D Bias Corrections}
\label{sec:BBC:subsec:Review}
A brief overview of the BBC process is described in Section~\ref{sec:Sims&Analysis:subsec:Analysis} and here we elaborate on the bias corrections component for BBC5D. \cite{Kessler16} expanded on the 1D method by defining a bias-corrected distance where the individual Tripp components of $m_B$, $c$, and $x_1$ are corrected. They define a bias-corrected distance, 
\begin{equation}
    \begin{aligned}
\mu^* &=   m_B^* + \alpha x_1^* - \beta c^*  - M_{\zindex}  \\ 
&=  (m_B - {\delta}_{m_B}) + \alpha (x_1 - {\delta}_{x_1}) - \beta (c - {\delta}_{c})  - M_{\zindex} \\
&= m_B + \alpha x_1 - \beta c  - M_{\zindex} - \delbias(z,x_1,c,\alpha,\beta) \label{eq:BCmu}
    \end{aligned}
\end{equation}
where bias-corrected quantities are denoted with a star superscript and 
\begin{equation}
    \delbias \equiv ({\delta}_{m_B} + \alpha {\delta}_{x_1} - \beta {\delta}_{c}) \label{eq:deldef}
\end{equation}
Measurement noise and intrinsic scatter preclude calculating the exact bias correction for each event; therefore, $\delbias$ is interpolated in 5D cells of $\{z, x_1,c,\alpha, \beta\}$. The $\delbias$ term is calculated with a large simulation designated as a `BiasCor', by comparing the observed values to the simulated ones for each SNIa parameter. The first three dimensions ($z$,$x_1$,$c$) are interpolated using the simulated populations. However, the luminosity correlation coefficients ($\alpha$, $\beta$) are single valued and not described with a population model akin to $z, x_1, c$. Therefore, the bias corrections for $\alpha$ and $\beta$ are determined on a $2\times2$ grid that brackets the values of $\alpha$ and $\beta$ found in the data. This grid enables for an interpolation of the BiasCor sample at the value of the proposed $\alpha$ and $\beta$ in each iteration of the BBC fit.

\subsection{Improving Bias Corrections to Account for SNIa-Host Correlations}
\label{sec:BBC:subsec:Ext}
\cite{Smith20} find that the fitted $\gamma$ from BBC5D is biased when analysing simulated samples that include correlations between $c / x_1$ and host-galaxy mass. To account for these biases in the mass step, we introduce two new dimensions to the $\delbias$ term in Equation \ref{eq:BCmu}: $\BiasCorStep$, which is a magnitude shift, and \mass.

The new $\BiasCorStep$ dimension is incorporated into the BiasCor by adding a magnitude shift of $ + \BiasCorStep$ to a random half of the simulation, and $ - \BiasCorStep$ to the other half; there is no correlation between $\BiasCorStep$ and host properties, and thus $\BiasCorStep \ne \gamma$. Using BiasCor with $\theta$, the BBC fit allows for a mag-shift ($\delhost$) as an arbitrary function of $X = $ $\log_{10}$\mass\ and other parameters. While previous cosmology analyses have used $\delhost = - \gamma / 2$ for $X > 10^{10}$ and $\delhost = + \gamma / 2$ for $X < 10^{10}$, here we adopt the more general $\delhost$ function with step location $S$ and step width $\tau_M$ (Equation \ref{eq:BCmu2}). At each step of the BBC fit, the value of $\delhost$ for each event is used to interpolate the BiasCor between $\pm \BiasCorStep$ such that
\begin{equation}
\begin{aligned}
& \dmuBiasCor = \label{eq:RICK} {\dmuBiasCor}(\vec{x}_5,X,-\BiasCorStep)  \\ 
& + \finterp \times   
     [ {\dmuBiasCor}(\vec{x}_5,X,+\BiasCorStep)  -
       {\dmuBiasCor}(\vec{x}_5,X,-\BiasCorStep) ] 
    \end{aligned}
\end{equation}
where $\finterp = (\delhost + \BiasCorStep) / 2\BiasCorStep $ and $\vec{x}_5 = \{z,x_1,c, \alpha, \beta \}$. It is important that $\theta > \delbias$ to ensure a valid interpolation.

The addition of $\theta$ and \mass\ results in changing the 5D $\delbias$ term in Equation \ref{eq:BCmu} to a 7D $\delbias$ as follows,

\begin{equation}
    \begin{aligned}
& \delbias (z, x_1,c,\alpha, \beta) \to \\ & \delbias (z, x_1,c,\alpha, \beta, \theta, M_{\textrm{stellar}})~.
\label{eq:The7D}
    \end{aligned}
\end{equation}

For the first six BiasCor dimensions, $\delbias$ is interpolated. For the \mass\ dimension, $\delbias$ is evaluated in discrete bins to avoid interpolating across the luminosity step at $10^{10} M_{\odot}$. 

\begin{figure}[h]
\includegraphics[width=9cm]{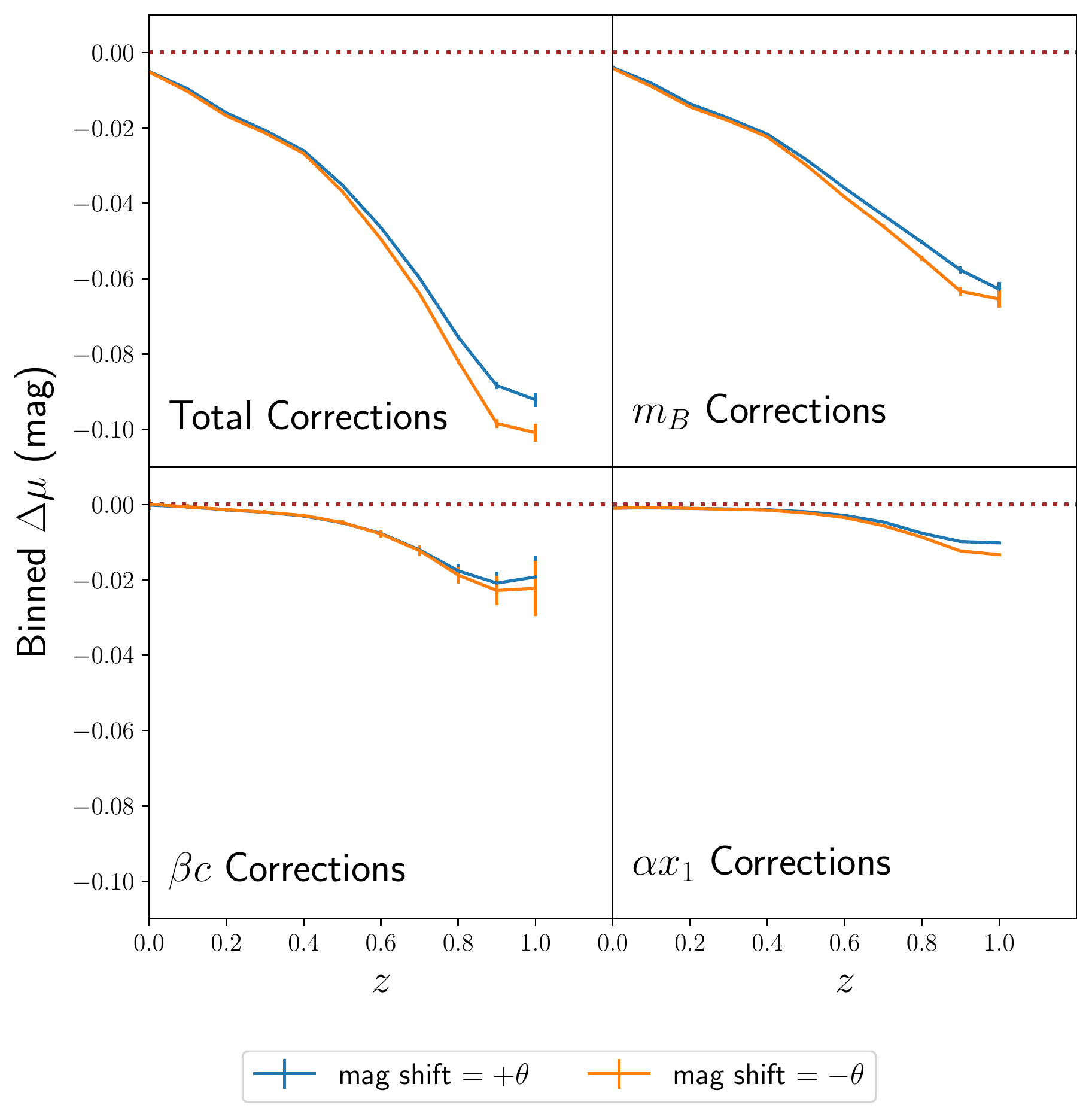}
\caption{Bias correction vs. redshift for $\theta = +0.06 $ (blue) and $\theta = -0.06$ (orange). The biased quantity is indicated on each panel.}
\label{fig:2GAMMAS}
\end{figure}

The effect of $\theta$ on the bias corrections is shown in Figure \ref{fig:2GAMMAS} as a function of redshift. Here we set $\theta = \pm 0.06$, the effects of which is negligible for $z<0.8$ and increases to $\sim 0.02$ mag at high-$z$. Because $\theta$ is independent of supernova parameters, it can be used for investigating correlations between any host galaxy property and SNIa luminosity, not just host-galaxy stellar mass.

\subsection{Changes to the BBC formalism for BS20}\label{sec:BBC:subsec:BS20}

While the SALT2 model is an accurate description of SNIa light curves, it is nonetheless an approximation that  ignores the difference between intrinsic color variation and dust. Previous SNIa cosmology analyses have used the same SALT2 model in both light curve fitting and BiasCor; this means that the BiasCor corrects for selection effects, but does not correct for biases in the SALT2 model. BS20 attempts to provide a more accurate light-curve model that can be used for the BiasCor, and here we update BBC to be compatible with BS20.

The BBC5D formalism is not compatible with BS20 for three reasons: 1) BBC intrinsic scatter is characterised by a single colour-independent $\sigma_{\textrm{int}}$ while BS20 uses a dust-dependent scatter that is poorly characterised by a single $\sigma_{\textrm{int}}$, 2) BBC assumes the single SALT2 colour-luminosity relation $\beta$, whereas BS20 uses distributions for intrinsic $\beta_{\rm SN}$ and dust $R_V$ (Eq. \ref{eq:BS20-deltamb}), and 3) $c_{\rm int}$ and $\beta_{\rm SN}$ in BS20 refer to intrinsic colour and intrinsic colour-luminosity relationship, while SALT2 $c$ and $\beta$ include both dust and intrinsic properties.

Ideally, to incorporate BS20, we would decompose the fitted colour $c$ into $c_{\rm int}$ and $c_{\rm dust}$ and compute a true $\beta$ in the simulation. While we plan to address these issues in a future work, they require significant updates to both training and fitting code; instead, we make several approximations. Here we describe BBC updates to be compatible with the BS20 model:

\begin{enumerate}
    \item Replace the SALT2 model in the BiasCor with the BS20 model.
    \item Include \mass\ as in Section~\ref{sec:BBC:subsec:Ext}, but not $\theta$ because the mass step is predicted by the BS20 model without the $\gamma$ parameter.
    \item Remove grid-interpolation of $\alpha$ and $\beta$ because the BS20 simulations are forward-modeled and result in BBC-fitted $\alpha$ and $\beta$ that are in agreement with that of the data.
    \item $\delbias$ is computed for distances instead of bias-correcting each SALT2 parameter.
\end{enumerate}
These changes are referred to as BBC-BS20 and result in a $\delbias$ dimensionality of $\{z, x_1, c, M_{\textrm{stellar}}\}$. For BBC-BS20, Equation \ref{eq:BCmu} becomes:

\begin{equation}
    \mu^* = m_B + \alpha x_1 - \beta c - M_{\zindex} - \delbias
    \label{eq:BS20BC1}
\end{equation}
and Equation~\ref{eq:deldef} becomes
\begin{equation}
    \delbias = m_B^{\textrm BC} + \alpha x_1^{\textrm BC} - \beta c^{\textrm BC} - M_{\zindex} - \mu_{\textrm{true}}~.
    \label{eq:BS20BC2}
\end{equation}
The true distance modulus for the BiasCor is $\mu_{\rm true}$, and the superscript BC denotes that these values are from SALT2 light-curve fits to the BiasCor events. We note that Equations~\ref{eq:BS20BC1} and \ref{eq:BS20BC2} can be used with BBC5D and the SALT2 model, and yield consistent results compared to using Equations \ref{eq:BCmu} and \ref{eq:deldef}. 

\section{Results}\label{sec:Results}

Here we evaluate the accuracy of our population modelling and bias correction approaches. Section~\ref{sec:Results:subsec:ParentPops} discusses the results of the parent population modelling and how well our simulations match the data. Section~\ref{sec:Results:subsec:BBC} presents a comparison of BBC7D to previous approaches for a variety of metrics. Section~\ref{sec:Results:subsec:BS20} contains the results of the new BBC-BS20 method for dust-scatter models. A discussion of these results is in Section \ref{sec:Conclusion}.

\begin{figure}
\includegraphics[width=9cm]{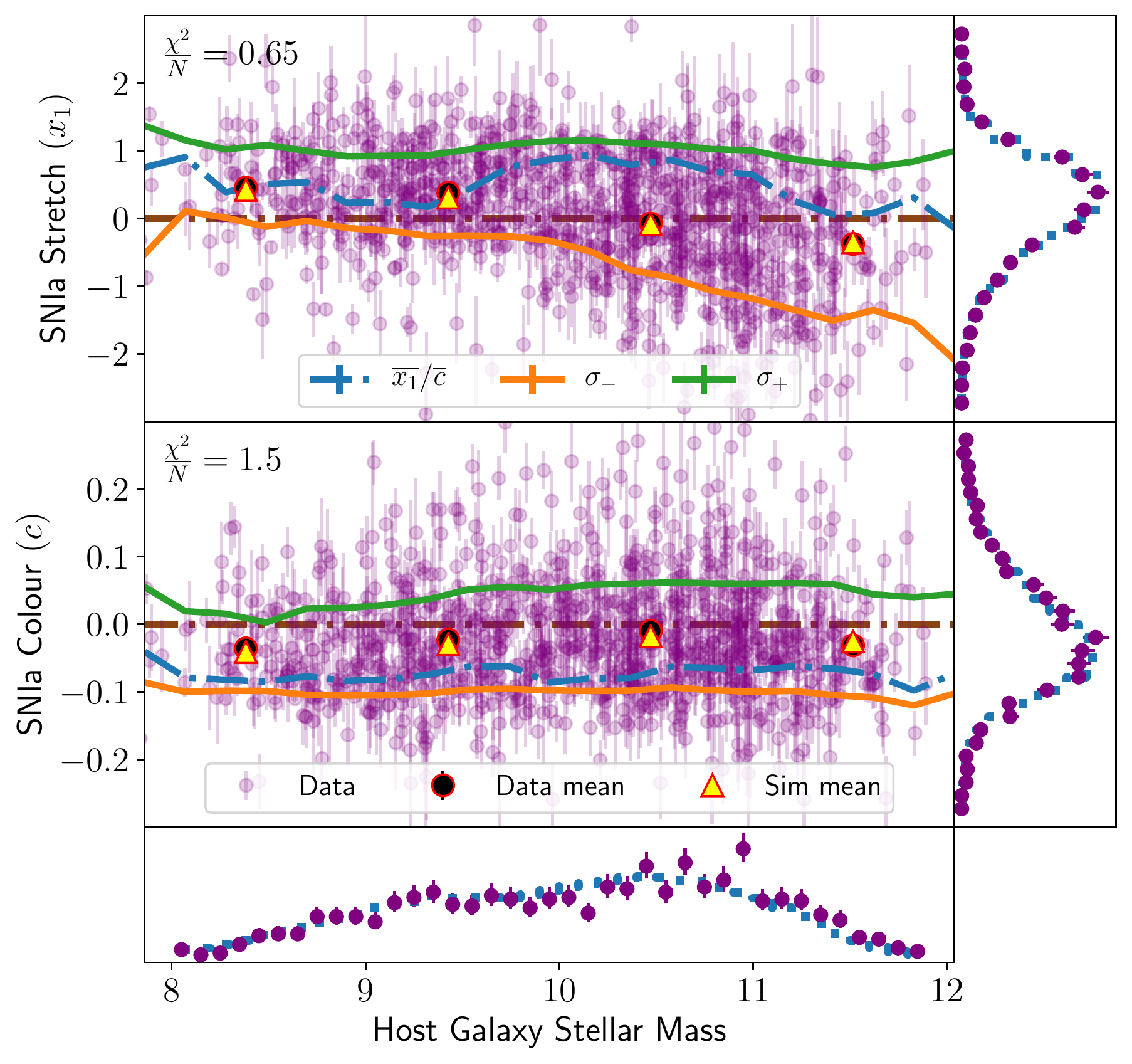}
\caption{The distributions of $c$ and $x_1$ for the combined DES, PS1, SNLS, and SDSS spectroscopic samples as a function of host-galaxy mass. Circles are the data; purple represent individual data points and black are the average value for that bin. Solid lines are the G10 parent population parameters: green is $\sigma_+$, blue is the peak value, and orange is $\sigma_-$. The dash-dotted line is 0. The bin averages for the data are presented in red-lined black dots, for the simulations, red-lined yellow triangles. The histograms on the side show data in circles and simulated results in dashed line. For both $c$ and $x_1$ we find good agreement between the mean value in data and sims, with a $\chi^2/N = 1.5$ and  $\chi^2/N = 0.65$ respectively. }
\label{fig:obs-dist-2}
\end{figure}

\begin{figure}
\includegraphics[width=9cm]{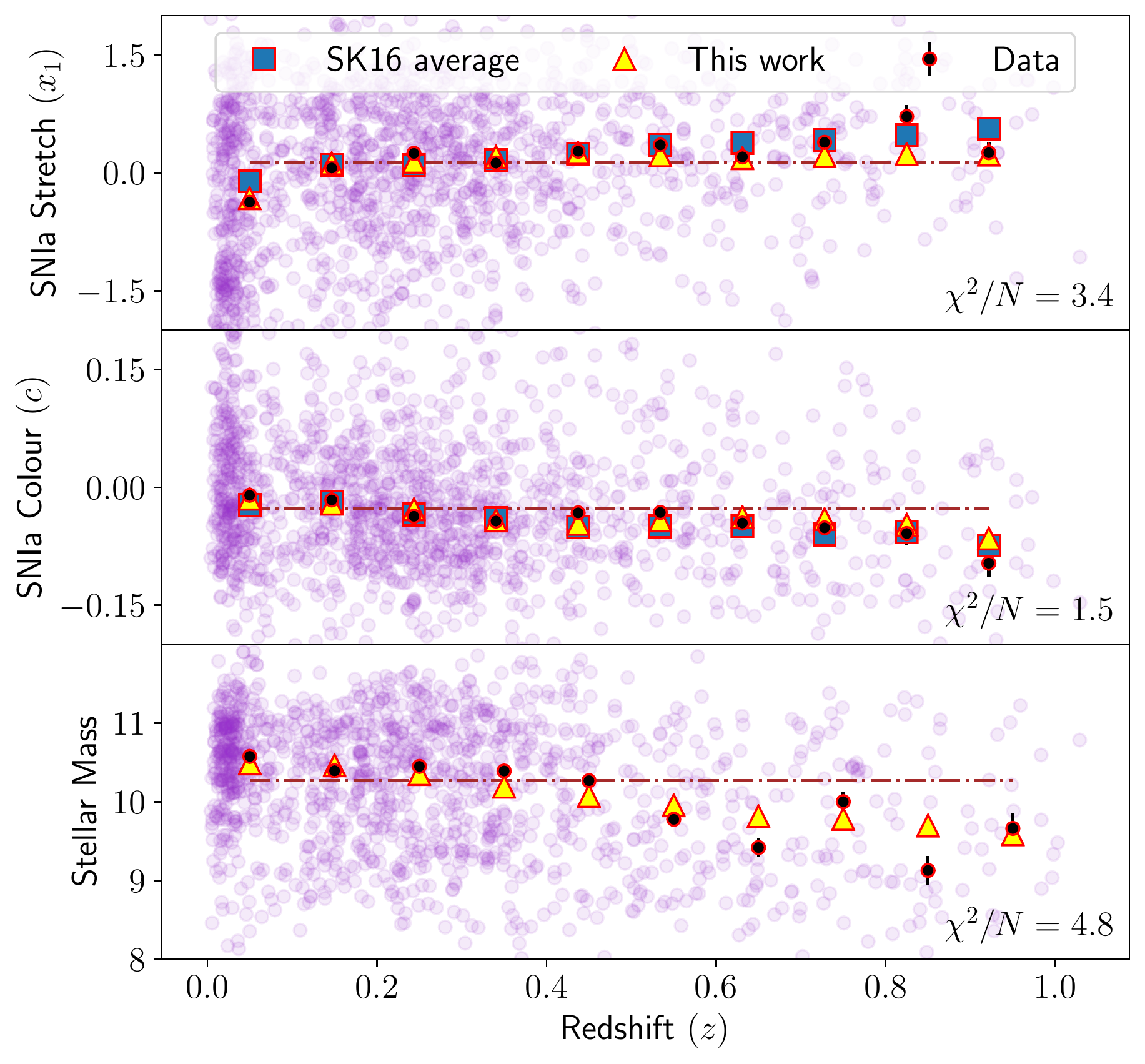}
\caption{The redshift evolution of $c$, $x_1$, and host-galaxy stellar mass for the combined sample. The data is presented in purple circles, and averaged bin values for the data are shown in red-lined black circles. The averaged bin values for the SK16 simulations are shown in red-lined blue squares, red-lined yellow triangles represent the average bin values for this work. Individual simulated events are not shown for clarity. We present the $\chi^2/N$ values for each parameter. SK16 assumes a flat distribution of \mass with redshift.}
\label{fig:ZDEP}
\end{figure}

\subsection{Parent Populations}\label{sec:Results:subsec:ParentPops}
Using the simulations described in Section~\ref{sec:Sims&Analysis:subsec:Sims}, we follow the process presented in Section~\ref{sec:Modeling:subsec:Extend} to determine the underlying populations of stretch and colour when assuming the G10 or C11 scatter models. For both scatter models, the population parameters for the generalised normal distribution, \sighigh, $\overline{x}_1$, \siglow, are presented in Appendix A2 for each individual survey. These population parameters are provided in steps of $0.2 \times 10^{10} M_{\textrm{stellar}}$. 

In Figure \ref{fig:obs-dist-2}, we show the comparison between sim and data for $c$ vs. \mass\ and for $x_1$ vs. \mass. We also show the \mass\ dependence of the underlying combined SDSS, DES, SNLS, and PS1 parent population. The Foundation and Low-z parent populations are shown separately in Figures \ref{fig:SURVEY-c} and \ref{fig:SURVEY-x1} for colour and stretch respectively. There are three notable results. First, there is excellent agreement between the observed mean values for data and simulation. In the combined SDSS, DES, SNLS, and PS1 sample, this parent population also characterises the constituent individual surveys. Second, we find that both the intrinsic $x_1$ and $c$ populations depend on \mass. 

We evaluate the significance of \mass\ dependence by comparing each curve in Figure \ref{fig:obs-dist-2} to a null model that has no \mass\ dependence. For the $x_1$ distribution, \sighigh~ is consistent with no \mass\ dependence with a confidence of 99.9\%. The probability that \siglow~ has no \mass\ dependence, however, is 1E-8. This means that the observed $x_1$ dependence on mass is driven by \siglow~ increasing with increasing masses. Similarly, $\overline{x_1}$ has a low probability of being independent of \mass~ at only 0.01\%, but does not correlate with the observed distribution and is therefore unlikely to be the primary driver. For the colour distribution, \siglow~ and $\overline{c}$ are consistent with no \mass\ dependence with a confidence of 90\%. Similarly to $x_1$, the probability that the faint-side $\sigma_{+}$ has no \mass\ dependence is 99.9\%. Taken together, this constitutes the third notable result: the \mass\ dependence of the observed $c$ and $x_1$ distributions is driven by increasing faint-side widths (\siglow~ for $x_1$ and \sighigh~ for $c$) rather than a shift in the mean values.

We also investigate the observed redshift dependence of our $c$ and $x_1$ distributions. Figure \ref{fig:ZDEP} shows that the observed colour and stretch distributions depend on redshift for two reasons: 1) separate populations for low and high $z$ (Section \ref{sec:Modeling}), and 2) selection effects. The data-simulation agreement is reasonable, with $\chi^2/N = $ 4.8, 1.5, and 3.5 for stretch, colour, and \mass, respectively.

\begin{figure}
\includegraphics[width=9cm]{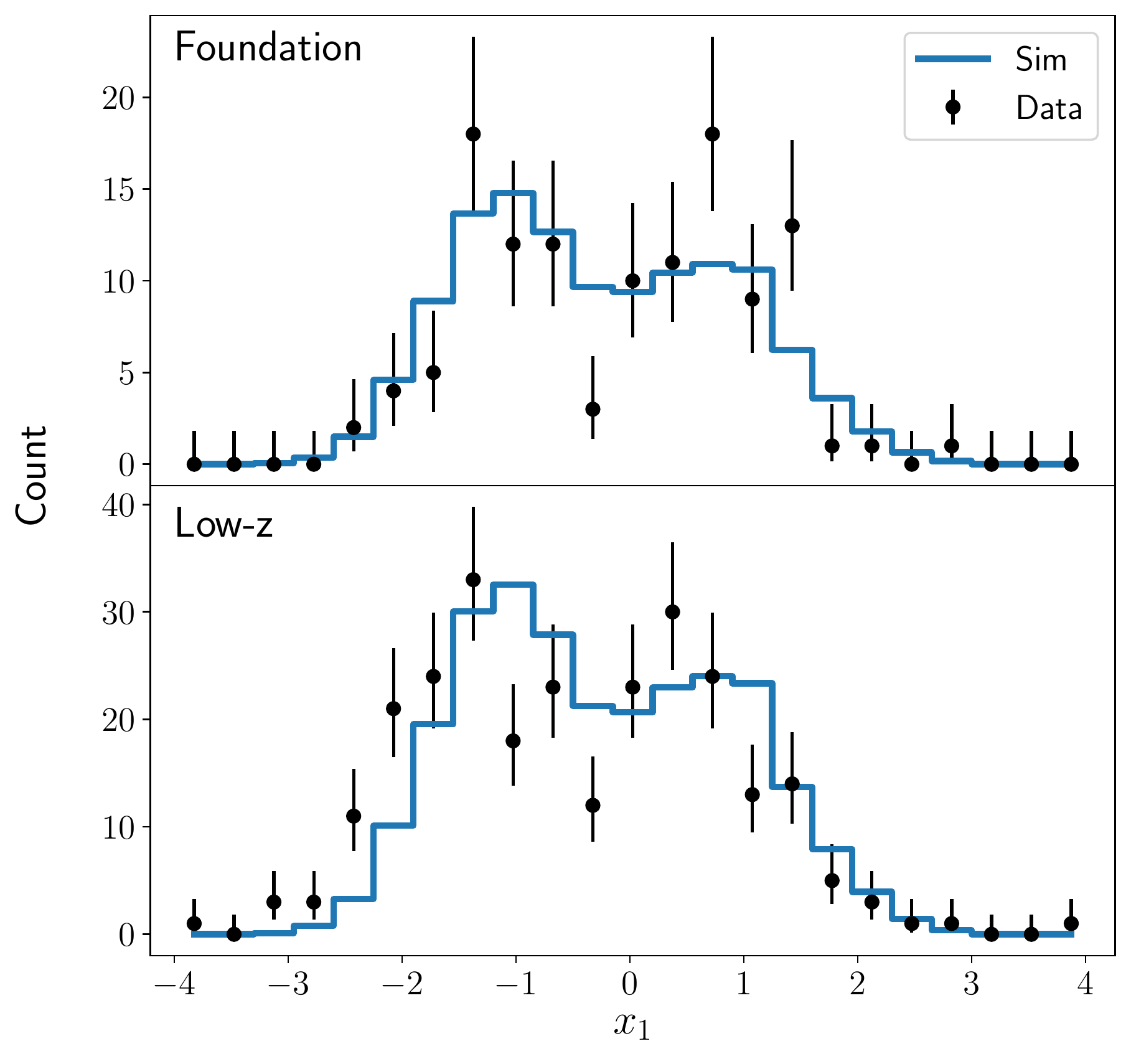}
\caption{The $x_1$ distributions for Foundation and Low-z surveys. Data is presented in black circles and simulation is presented in blue histogram.}
\label{fig:low-redshift-x1}
\end{figure}

To model the double-peaked $x_1$ distribution in Low-z and Foundation, we found it necessary to implement a prior requiring that $\overline{x_1}_1 < 0$ and $\overline{x_1}_2 > 0$ (Eq. \ref{eq:DGauss}). While we are able to capture the double-peak behaviour for both surveys, shown in Figure \ref{fig:low-redshift-x1}, there is a slight data-sim discrepancy near $x_1 \sim -2$ for the Low-z sample.

\subsection{Impact on Cosmology Using Bias Corrections with Host Properties}\label{sec:Results:subsec:BBC}

With our simulations from Section~\ref{sec:Sims&Analysis:subsec:Sims} and the parent populations shown in Figure \ref{fig:obs-dist-2}, we create 100 simulated data sets and a large BiasCor to test the accuracy of our bias correction methods. We compare four different bias correction methods for the G10 and C11 models. Two methods use BBC1D, with and without \mass\ information included in the BiasCor (designated `No Mass', similar to that of \cite{Betoule14}). The other two methods use the BBC5D and new BBC7D method. Overall, we find our new method significantly improves upon BBC5D. In Table \ref{tab:BigTable}, we show the fitted nuisance parameters, Hubble scatter (RMS of $\mu_{\rm fit} - \mu_{\rm true}$), $w$ and \wbias\ defined as $\Delta w = (w_{\rm fit} - w_{\rm true})$, averaged over the 100 simulations. 

For the G10 scatter model, we find an $\alpha$ bias below $1\%$ for all BBC methods. This broad agreement holds well for the recovery of $\beta$ as well. The BBC5D approach results in a significant $\gamma$ bias of $\Delta \gamma = 0.019 \pm 0.001$~mag. BBC7D reduces this bias by a factor of $\sim 5$, although a 2 millimag $\gamma$-bias remains with $2 \sigma$ significance. These trends in $\alpha$, $\beta$, and $\gamma$ are qualitatively similar in the C11 results. However, it is worth noting that the BBC1D recovery of $\beta$ and $\gamma$ for the C11 scatter model is significantly biased.

For both G10 and C11, we see a general decrease in Hubble scatter with increasing dimensionality of the BBC methodology. BBC5D and BBC7D have comparable Hubble scatter with each other for both G10 and C11, and both are $\sim 10\%$ smaller compared to using BBC1D. 

The presence of host-galaxy correlations in the BBC1D BiasCor does not make a significant impact on recovered $w$ values for G10 or C11. In the case with and without host-galaxy correlations, the \wbias~ is $\sim 0.02 \pm 0.005$ for G10 and $\sim 0.01 \pm 0.005$ for C11. BBC5D does not significantly effect this BBC1D \wbias, recovering $- 0.0201 \pm 0.0048$ and $-0.0237 \pm 0.0046$ for G10 and C11 respectively. The BBC7D \wbias ~ for both G10 and C11 is a factor of 2 smaller than their 5D counterparts:
\wbias $= 0.0086 \pm 0.0046$ for G10 and ($0.0105 \pm 0.0046$ for C11. Both BBC7D biases have $\sim 2\sigma$ significance. Overall, BBC7D has the smallest \wbias~ as well as the smallest biases for other nuisance parameters and scatter.

\begin{figure}[t]
\includegraphics[width=9cm]{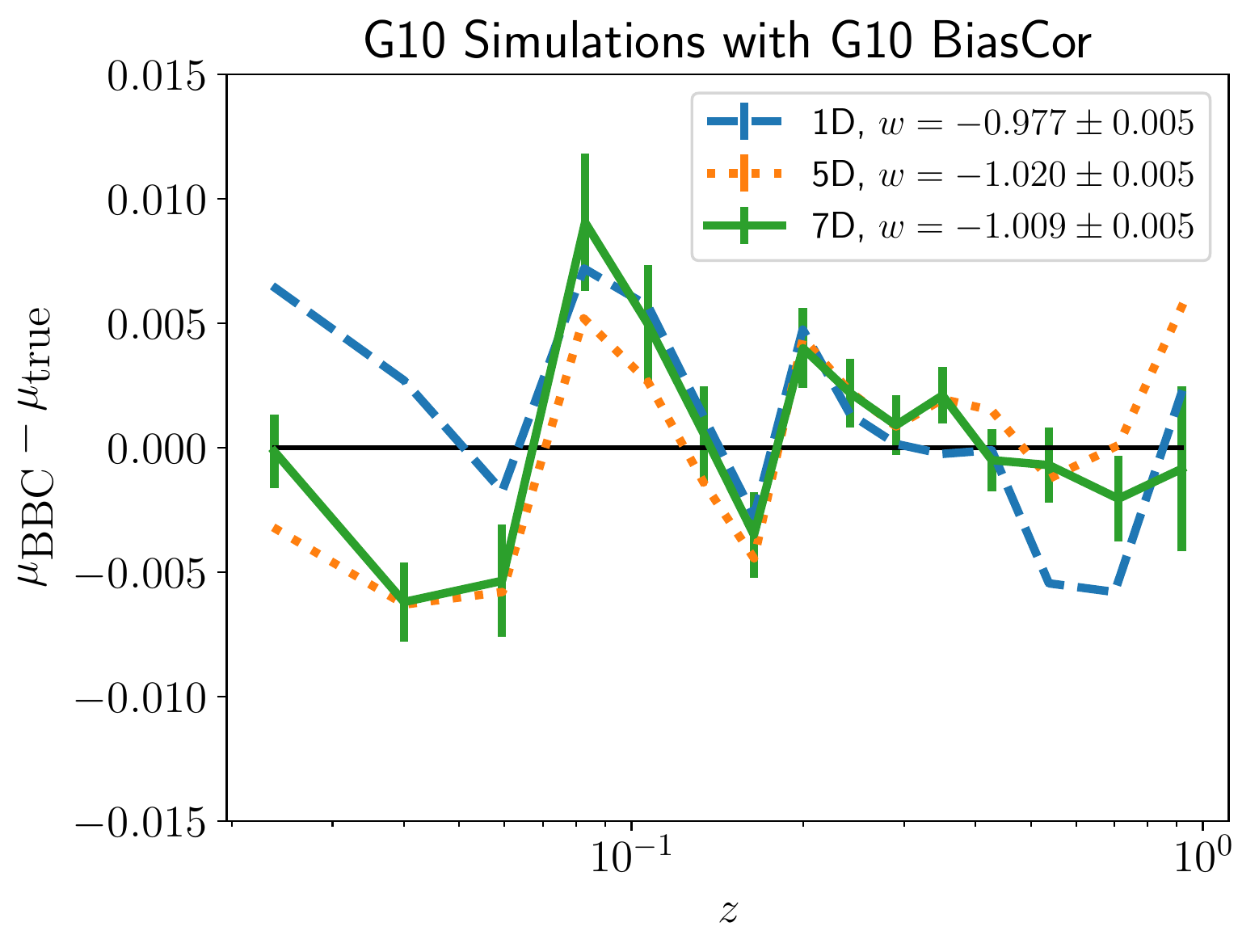}
\caption{The binned $\mu$ residuals for the G10 scatter model using three different BBC methods on the same set of simulations. Uncertainties are shown for BBC7D only but are representative for the other methods.}
\label{fig:Corr-M0DIFF}
\end{figure}

In Figure \ref{fig:Corr-M0DIFF}, we show the binned distance modulus residuals as a function of redshift for BBC1D, BBC5D, and BBC7D. BBC7D has smaller $\mu$ residuals (mostly $\sim 0.005$ across the $z$ range) than the BBC5D approach. All BBC $\mu$ residuals have a significant excess, $\mu_{\rm BBC} - \mu_{\rm true} = 0.01 \pm 0.004$ ($2.4\sigma$), around $z = 0.1$.

\begin{figure}
\includegraphics[width=8cm]{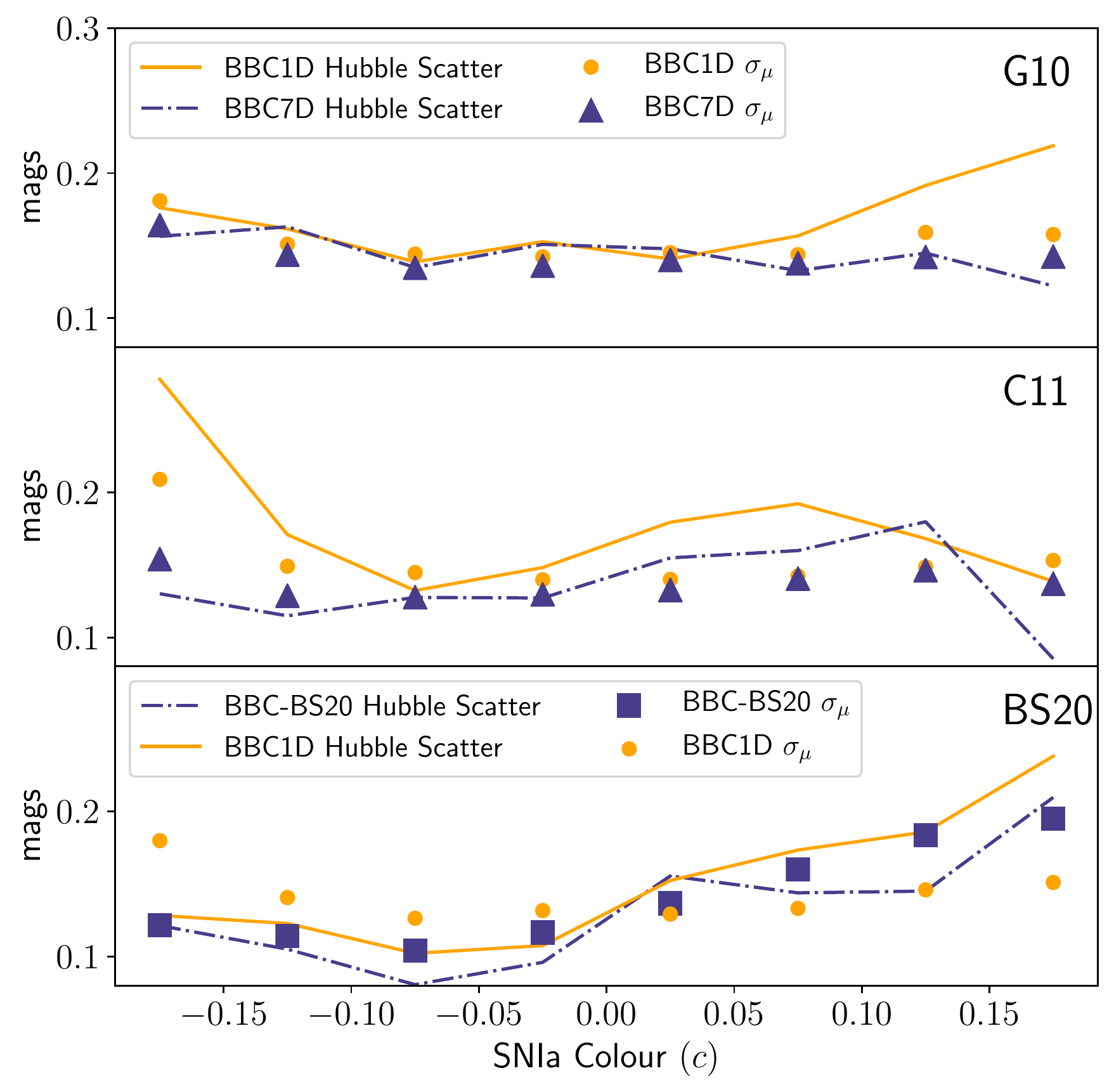}
\caption{The Hubble Scatter (line) and the BBC-fitted distance modulus uncertainties, $\sigma_{\mu}$, (triangle) are plotted for three different scatter models. We expect good agreement between the two for an effective BBC method.}
\label{fig:Cov}
\end{figure}

The accuracy of the reported distance modulus error, $\sigma_{\mu}$, in comparison to Hubble Residual (HR) scatter is another important metric in determining the effectiveness of BBC. Figure \ref{fig:Cov} shows BBC1D and BBC7D $\sigma_{\mu}$ and HR scatter as a function of SNIa colour, $c$. We see better agreement for BBC7D compared to BBC1D. Figure \ref{fig:Cov} and Table \ref{tab:BigTable} show that for G10 and C11, the BBC7D methods have the smallest scatter.

\begin{table*}[t]
\scalebox{1}{%
\begin{tabular}{cccccccc}
    Method & Model & input $\alpha = 0.145$ & input $\beta = 3.1$ & input $\gamma = 0.05$ &  & input $w = -1$ & \\
    \hline
    & & Fitted $\alpha$ & Fitted $\beta$ & Fitted $\gamma$  & Hubble scatter \footnotemark[2] & Fitted $w$ & $\Delta_w ~ (N_{\sigma_w})$ \\ 
    \hline
    BBC1D (No Mass) \footnotemark[3] & G10 & \ddqa{} & \ddqb{} & \ddqg{} & \ddqst{} & \ddqw{} & $~0.017 ~ (3.7 \sigma)$\\
    BBC1D & G10 & \ddaa{} & \ddab{} & \ddag{} & \ddast{} & \ddaw{} & $~0.022 ~ (4.7 \sigma)$\\
    BBC5D & G10 & \ddfa{} & \ddfb{} & \ddfg{} & \ddfst{} & \ddfw{} & $-0.0201 ~ (4.4 \sigma)$\\
    BBC7D & G10 & \ddca{} & \ddcb{} & \ddcg{} & \ddcst{} & \ddcw{} & $-0.0086 ~ (1.9 \sigma)$\\
     & & & & & \\
    Method & Model & input $\alpha = 0.145$ & input $\beta = 3.8$ & input $\gamma = 0.05$ &  & input $w = -1$ & \\
    \hline
     & & Fitted $\alpha$ & Fitted $\beta$  & Fitted $\gamma$  & Hubble scatter & Fitted $w$ & $\Delta_w ~ (N_{\sigma_w})$ \\ 
    \hline
    BBC1D (No Mass) & C11 & \bbqa{} & \bbqb{} & \bbqg{} & \bbqst{} & \bbqw{} & $~0.0104 ~ (2.6 \sigma)$\\
    BBC1D & C11 & \bbaa{} & \bbab{} & \bbag{} & \bbast{} & \bbaw{} & $~0.0039 ~ (0.8 \sigma)$\\
    BBC5D & C11 & \bbfa{} & \bbfb{} & \bbfg{} & \bbfst{} & \bbfw{} & $-0.0237 ~ (5.2 \sigma)$\\
    BBC7D & C11 & \bbca{} & \bbcb{} & \bbcg{} & \bbcst{} & \bbcw{} & $-0.0105 ~ (2.3 \sigma)$\\
     & & & & & \\
    Method & Model & input $\alpha = 0.145$ & input $\beta = \textrm{ND}$ \footnotemark[4] & input $\gamma = 0$ &  & input $w = -1$ &  \\
    \hline
     & & Fitted $\alpha$ & Fitted $\beta$ & Fitted $\gamma$  & Hubble scatter & Fitted $w$ & $\Delta_w ~ (N_{\sigma_w})$ \\ 
    \hline
    BBC1D (No Mass) & BS20 & \eeqa{} & \eeqb{} & \eeqg{} & \eeqst{} & \eeqw{} & $~0.0455 ~ (10.1 \sigma)$ \\
    BBC1D & BS20 & \eeaa{} & \eeab{} & \eeag{} & \eeast{} & \eeaw{} & $~0.0428 ~ (9.5 \sigma)$ \\
    BBC-BS20 & BS20 & \eena{} & \eenb{} & \eeng{} & \eenst{} & \eenw{} & $~0.0057 ~ (1.4 \sigma)$ \\
    \textcolor{white}{space} & \textcolor{white}{space} & \textcolor{white}{space} & \textcolor{white}{space} & \textcolor{white}{space} & \textcolor{white}{space} & \textcolor{white}{space} \\
    \footnotetext[1]{Uncertainties are the average uncertainty among 100 simulations, divided by $\sqrt{100}$} \footnotetext[2]{The error in the Hubble scatter is the R.M.S divided by $\sqrt{100}$.} \footnotetext[3]{Does not include \mass\ dependent parent populations in the BiasCor.} \footnotetext[4]{Not Determined: For the BS20 model, input SALT2 $\beta$ $(\beta \neq \beta_{\rm SN})$ is not defined.} 
\end{tabular}%
}
\caption{Fitted values and uncertainties \protect\footnotemark[1] averaged over 100 simulations \label{tab:BigTable}}
\end{table*}

\subsection{Impact on Cosmology Using Bias Corrections for BS20}\label{sec:Results:subsec:BS20}

The bottom tier of Table \ref{tab:BigTable} shows how well BBC1D and BBC-BS20 perform on simulations generated with the BS20 model. While simulations of the BS20 model do not have an input $\gamma$ like the aforementioned simulations G10 and C11, the simulation includes host-mass vs colour-luminosity dependence that matches that of the data. As reported in BS20, we find that using BBC1D results in $\gamma \sim 0.06$, 
consistent with the value found in the data. 
BBC-BS20 results in $\gamma = 0.0006 \pm 0.0007$, consistent with no mass step.  Applying BBC1D to BS20 simulations performs well in recovering $\alpha$ ($\Delta {\alpha}{\sim} 0.001 \pm 0.0006$),  while BBC-BS20 has a slightly larger $\alpha$-bias ($\Delta {\alpha} \sim 0.003 \pm 0.0004$). To validate the predicted $\gamma$-reduction, we apply the BBC-BS20 method on real data and find
\begin{equation}
    \gamma = 0.0166 \pm 0.0076
\end{equation}
which is more than $\times 3$ smaller than previous measurements. 

The BS20 model is characterised by a distribution of $\beta_{\rm SN}$ and $R_V$, and thus we cannot determine an input $\beta$ for the simulation. For this reason, the BS20 section of Table \ref{tab:BigTable} refers to the input $\beta$ as Not Defined (ND), and we are unable to compute $\beta$ biases. BS20 found their model parameters such that their BBC1D $\beta$ from simulations matched that of the data ($\beta \sim 3.06 \pm 0.06$). Since BS20 used an overly simplistic assumption about \mass\ distributions, their fitted model parameters are approximate. Here we update the \mass\ distribution to match the data (Figure \ref{fig:obs-dist}) but do not update the BS20 model parameters; we find BBC1D correction on BS20 simulations results in $\beta \sim 2.76 \pm 0.01$, which no longer matches that of the data. However, when running BBC-BS20 on our simulations and real data we find good agreement: $\beta = 3.077 \pm 0.005$ (sim) and $\beta = 3.165 \pm 0.045$ (data) respectively. This BBC1D $\beta$ discrepancy warrants the need for further improvements to the BS20 modeling. 

BBC1D results in a \wbias~ of $\sim 0.04 \pm 0.004$, while the BBC-BS20 \wbias~is significantly reduced and consistent with 0, $\sim 0.006 \pm 0.004$. We find that the Hubble Scatter in the BS20 simulations is $\sim 10\%$ smaller than that of G10 or C11, which is illustrated in Figure \ref{fig:Cov}.

\section{Discussion and Conclusion}\label{sec:Conclusion} 
The utilisation of host galaxy information in SNIa standardisation analyses has become common in recent analyses. Typically, host galaxy properties have been included in the one of the final stages of the analysis to make an additive correction to SNIa luminosities before measuring cosmological parameters from the set of SNIa distances. However, this approach is conceptually flawed because it does not account for subtle biases due to the correlations of SNIa properties and host galaxies. Here, we have determined the underlying populations of SNIa properties and their correlations with host galaxies so that we can trace and correct for these biases.

Our approach for these corrections is to modify the BBC method to allow for higher dimensional bias corrections. If the intrinsic scatter model is known, BBC determines the input mass step ($\gamma$) with a bias of 0.004 mag for all models, and determines $w$ with biases consistent with 0 for G10 and BS20. For $\gamma$, the BBC7D bias is a factor of 5 smaller than for BBC5D; for $w$, the BBC7D bias is a factor of 2 smaller.

We do not address potential biases if the intrinsic scatter model is not known. Using BBC5D, \cite{Scolnic18} and \cite{Brout19b} showed that the difference in $w$ between assuming the G10 and C11 is $\sim 0.03$. Using BBC1D, \cite{BS20} showed that the $w$-difference when doing 1D corrections between BS20, G10, and C11 was as much as much as 4\%. However, in that study, they did not simultaneously refit for a mass step, and it is unclear how this approximation would affect the cosmological bias. This  intrinsic scatter uncertainty will be quantified in an upcoming analysis (Popovic et al. in prep.), which will determine optimal parameters of the BS20 model and address the discrepancy in the fitted 1D $\beta$ values between data and simulation. 

In determining the underlying populations, we follow a purely empirical approach that does not rely on theoretical models to relate SNIa properties to host galaxy properties. \cite{Rigault2020} use theoretical modeling to predict an evolution of the relation between the $x_1$ distribution with redshift due to the different sampling of progenitor systems at high-$z$ compared to low-$z$. While we are unable to show whether this evolution is due to progenitor systematics or survey selection effects, our analysis illustrates how the model from \cite{Rigault2020} can be incorporated in a cosmological analysis.

From our empirical approach, we find that peak probabilities of the parent distributions of color and stretch do not evolve significantly with host-galaxy stellar mass, but rather that the observed dependence on \mass~ is explained by increasing asymmetries of the parent distributions. For $c$, this finding agrees with the model proposed in BS20, who suggested that dust is responsible for one-sided SNIa $c$ scatter towards redder colors. Similarly, \cite{Rigault18} propose that local specific star formation rate (lsSFR) is a tracer of the progenitor age which itself traces the SNIa $x_1$ distribution. They find that the parent distributions of $x_1$ for low lsSFR and high lsSFR are inconsistent. We find that the observed correlation between $x_1$ and mass is driven by an increasing faint side standard deviation ($\sigma_-$) of the parent distribution, which suggests a more subtle relationship between $x_1$ and mass that can be described by a non-evolving peak probability and an evolving asymmetry.

Finally,  while we focused here on using host-galaxy mass to correlate with SNIa properties, our methods can be applied to other host galaxy properties. Furthermore, while we used a step function for the mass dependence, an arbitrary functional form can be used with BBC7D. Ultimately, our improved BBC method will be evaluated by varying
the SNIa model such that simulated distributions match those of the data, and propagating these variations to the BiasCor used by BBC.

\subsection{Acknowledgements}
Dillon Brout acknowledges support for this work was provided by NASA through the NASA Hubble Fellowship grant HST-HF2-51430.001 awarded by the Space Telescope Science Institute, which is operated by Association of Universities for Research in Astronomy, Inc., for NASA, under contract NAS5-26555. 

Richard Kessler acknowledges that this work was supported in part by the Kavli Institute for Cosmological Physics at the University of Chicago through an endowment from the Kavli Foundation and its founder Fred Kavli. R.K. is supported by DOE grant DE-SC0009924. Dan Scolnic is supported by DOE grant DE-SC0010007 and the David
and Lucile Packard Foundation.

\newpage

\appendix

\subsection{A1. EFFICIENCIES AND ERRATA}
\begin{figure}
\includegraphics[width=8cm]{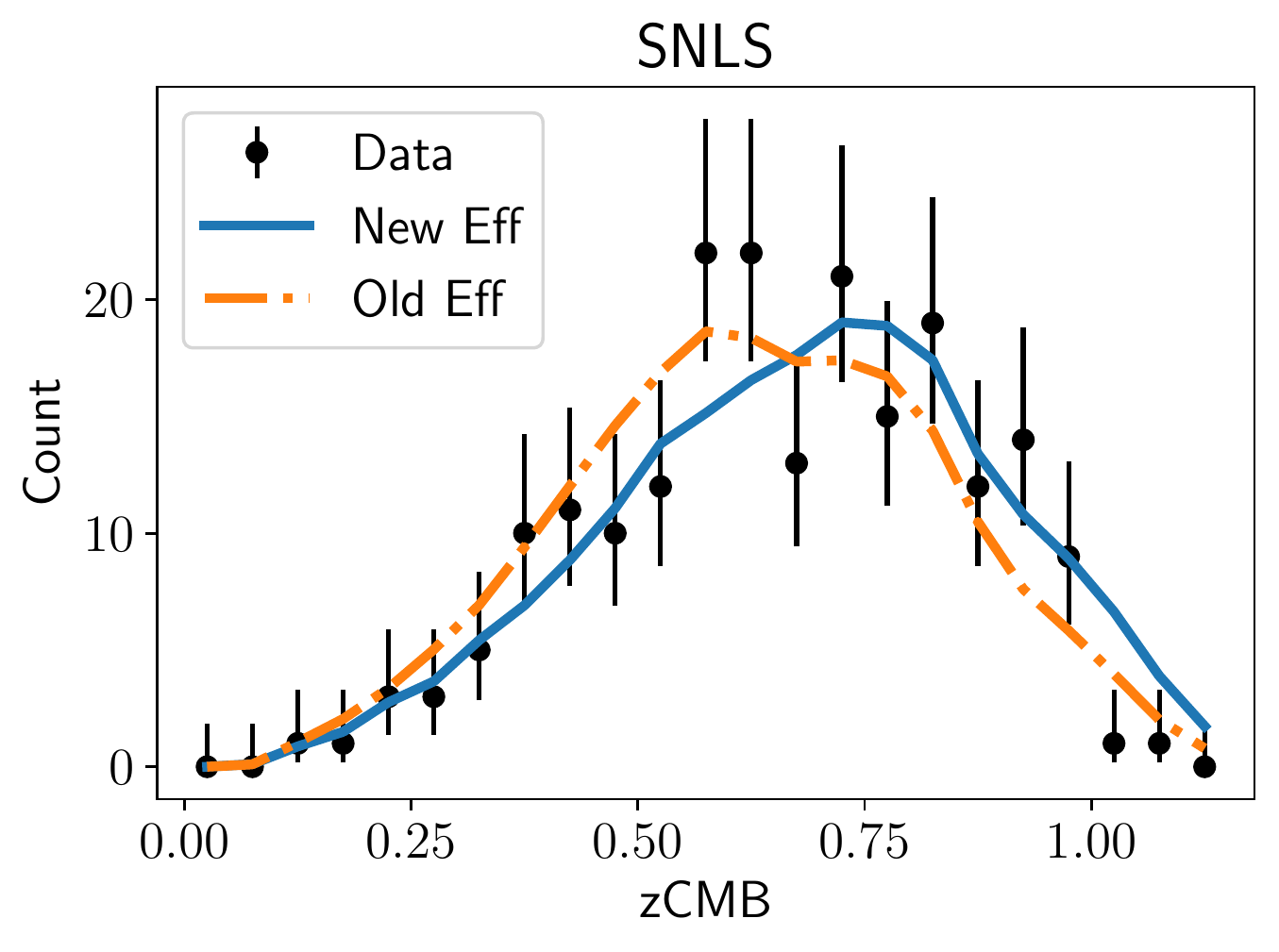}
\caption{The SNLS redshift distribution for the data (solid black line with points), simulations with JLA spectroscopic efficiency (orange dash-dot line), and updated spectroscopic efficiency presented here (blue solid line). At higher redshifts, the new efficiency better matches the data.}
\label{fig:SNLS_EFF}
\end{figure}

For the SNLS sample, we improve the spectroscopic efficiency function described as a function of peak $i$-band mag, Eff$_{\textrm{spec}(i)}$. Figure \ref{fig:SNLS_EFF} shows that the JLA Eff$_{\textrm{spec}}$, also used in BS20, underestimates the number of supernovae at redshifts greater than $0.75$. Our update is to replace Eff$_{\textrm{spec}(i)}$ with Eff$_{\textrm{spec}(i-0.3)}$.

\subsection{A2. PARENT POPULATION PARAMETERS}

\begin{figure*}[h!]
\includegraphics[width=18cm]{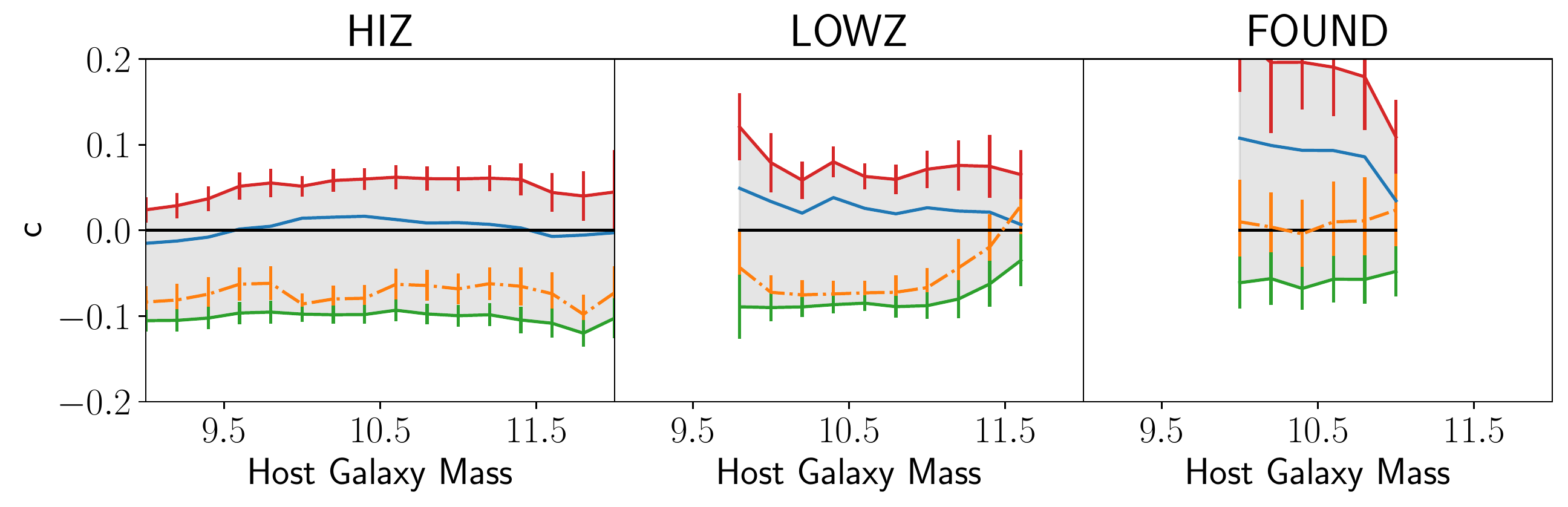}
 \caption{The G10 $c$ parent populations as a function of host-galaxy mass for several spectroscopic surveys. The mean of the asymmetric distribution is presented in blue, the peak in orange, the bright-side standard deviation in green, and the faint-side standard deviation in red. Errors are included for each parameter and the 68\% confidence interval for each is shown in grey fill. }
\label{fig:SURVEY-c}
\end{figure*}

\begin{figure*}[h!]
\includegraphics[width=18cm]{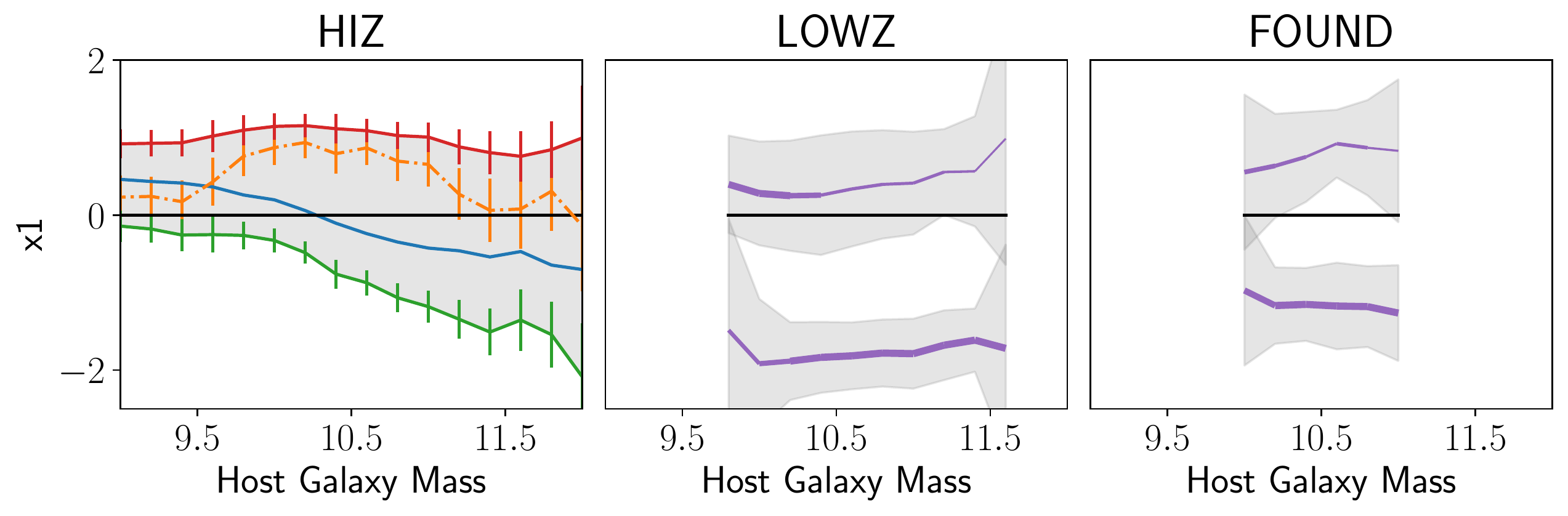}
\caption{The G10 $x_1$ parent populations as a function of host-galaxy mass for several spectroscopic surveys. The mean of the asymmetric distribution is presented in blue, the peak in orange, the faint-side standard deviation in green, and the bright-side standard deviation in red. Errors are included for each parameter and the 68\% confidence interval for each is shown in grey fill. For Foundation and Low-z, two peak values are shown by purple lines, and the line thickness denotes the relative weight of each peak.}
\label{fig:SURVEY-x1}
\end{figure*}

Here we show the parent populations as a function of mass for each survey, characterised by an asymmetric generalised normal distribution with three parameters: a peak probability and two standard deviations. See Equation \ref{eq:AGauss} for rigorous definition. For $c$, the $n$ value is fixed at $n=2$. For $x_1$, $n=3$. For the Low-z and Foundation surveys, we present a double-Gaussian approach characterised in Equation \ref{eq:DGauss}.
Population fit results for $c$ are in Tables \ref{tab:LOWZ_parents_c}, \ref{tab:FOUND_parents_c}, \ref{tab:COMB_parents_c} for Low-z, Foundation, and the combined DES, SDSS, SNLS, and PS1 sample, respectively. The population fit results for $x_1$ are in Tables \ref{tab:LOWZ_parents_x1} \ref{tab:FOUND_parents_x1}, \ref{tab:COMB_parents_x1} for Low-z, Foundation, and the combined DES, SDSS, SNLS, and PS1 sample, respectively. These populations are shown in Figures \ref{fig:SURVEY-c} and \ref{fig:SURVEY-x1} for $c$ and $x_1$ for the G10 scatter model. 

\begin{table}
    \centering
    \begin{tabular}{ccccccc}
        Mass & mean (G10) & $\sigma_-$ (G10) & $\sigma_+$ (G10) & mean (C11) & $\sigma_-$ (C11) & $\sigma_+$ (C11) \\ 
        \hline
9.8 & \LOWZcjmean{} & \LOWZcjstdl{} & \LOWZcjstdr{} & \LOWZcCjmean{} & \LOWZcCjstdl{} & \LOWZcCjstdr{} \\
10.0 & \LOWZckmean{} & \LOWZckstdl{} & \LOWZckstdr{} & \LOWZcCkmean{} & \LOWZcCkstdl{} & \LOWZcCkstdr{} \\
10.2 & \LOWZclmean{} & \LOWZclstdl{} & \LOWZclstdr{} & \LOWZcClmean{} & \LOWZcClstdl{} & \LOWZcClstdr{} \\
10.4 & \LOWZcmmean{} & \LOWZcmstdl{} & \LOWZcmstdr{} & \LOWZcCmmean{} & \LOWZcCmstdl{} & \LOWZcCmstdr{} \\
10.6 & \LOWZcnmean{} & \LOWZcnstdl{} & \LOWZcnstdr{} & \LOWZcCnmean{} & \LOWZcCnstdl{} & \LOWZcCnstdr{} \\
10.8 & \LOWZcomean{} & \LOWZcostdl{} & \LOWZcostdr{} & \LOWZcComean{} & \LOWZcCostdl{} & \LOWZcCostdr{} \\
11.0 & \LOWZcpmean{} & \LOWZcpstdl{} & \LOWZcpstdr{} & \LOWZcCpmean{} & \LOWZcCpstdl{} & \LOWZcCpstdr{} \\
11.2 & \LOWZcqmean{} & \LOWZcqstdl{} & \LOWZcqstdr{} & \LOWZcCqmean{} & \LOWZcCqstdl{} & \LOWZcCqstdr{} \\
11.4 & \LOWZcrmean{} & \LOWZcrstdl{} & \LOWZcrstdr{} & \LOWZcCrmean{} & \LOWZcCrstdl{} & \LOWZcCrstdr{} \\
11.6 & \LOWZcsmean{} & \LOWZcsstdl{} & \LOWZcsstdr{} & \LOWZcCsmean{} & \LOWZcCsstdl{} & \LOWZcCsstdr{} 
    \end{tabular}
    \caption{LOWZ parent colour populations}
    \label{tab:LOWZ_parents_c}
\end{table}

\begin{table}
    \centering
    \begin{tabular}{ccccccc}
        Mass & Weight (G10) & $\overline{x}_1$ (G10) & $\sigma$ (G10) & Weight (C11) & $\overline{x}_1$ (C11) & $\sigma$ (C11) \\ 
        \hline
9.8 & \LOWZxjweightone{} & \LOWZxjmeanone{} & \LOWZxjstdone{} & \LOWZxCjweighttwo{} & \LOWZxCjmeantwo{} & \LOWZxCjstdtwo{} \\
10.0 & \LOWZxkweightone{} & \LOWZxkmeanone{} & \LOWZxkstdone{} & \LOWZxCkweighttwo{} & \LOWZxCkmeantwo{} & \LOWZxCkstdtwo{} \\
10.2 & \LOWZxlweightone{} & \LOWZxlmeanone{} & \LOWZxlstdone{} & \LOWZxClweighttwo{} & \LOWZxClmeantwo{} & \LOWZxClstdtwo{} \\
10.4 & \LOWZxmweightone{} & \LOWZxmmeanone{} & \LOWZxmstdone{} & \LOWZxCmweighttwo{} & \LOWZxCmmeantwo{} & \LOWZxCmstdtwo{} \\
10.6 & \LOWZxnweightone{} & \LOWZxnmeanone{} & \LOWZxnstdone{} & \LOWZxCnweighttwo{} & \LOWZxCnmeantwo{} & \LOWZxCnstdtwo{} \\
10.8 & \LOWZxoweightone{} & \LOWZxomeanone{} & \LOWZxostdone{} & \LOWZxCoweighttwo{} & \LOWZxComeantwo{} & \LOWZxCostdtwo{} \\
11.0 & \LOWZxpweightone{} & \LOWZxpmeanone{} & \LOWZxpstdone{} & \LOWZxCpweighttwo{} & \LOWZxCpmeantwo{} & \LOWZxCpstdtwo{} \\
11.2 & \LOWZxqweightone{} & \LOWZxqmeanone{} & \LOWZxqstdone{} & \LOWZxCqweighttwo{} & \LOWZxCqmeantwo{} & \LOWZxCqstdtwo{} \\
11.4 & \LOWZxrweightone{} & \LOWZxrmeanone{} & \LOWZxrstdone{} & \LOWZxCrweighttwo{} & \LOWZxCrmeantwo{} & \LOWZxCrstdtwo{} \\
11.6 & \LOWZxsweightone{} & \LOWZxsmeanone{} & \LOWZxsstdone{} & \LOWZxCsweighttwo{} & \LOWZxCsmeantwo{} & \LOWZxCsstdtwo{} \\
    \end{tabular}
    \caption{LOWZ parent stretch populations}
    \label{tab:LOWZ_parents_x1}
\end{table}


\begin{table}
    \centering
    \begin{tabular}{ccccccc}
        Mass & mean (G10) & $\sigma_-$ (G10) & $\sigma_+$ (G10) & mean (C11) & $\sigma_-$ (C11) & $\sigma_+$ (C11) \\ 
        \hline
10.0 & \FOUNDckmean{} & \FOUNDckstdl{} & \FOUNDckstdr{} & \FOUNDcCkmean{} & \FOUNDcCkstdl{} & \FOUNDcCkstdr{} \\
10.2 & \FOUNDclmean{} & \FOUNDclstdl{} & \FOUNDclstdr{} & \FOUNDcClmean{} & \FOUNDcClstdl{} & \FOUNDcClstdr{} \\
10.4 & \FOUNDcmmean{} & \FOUNDcmstdl{} & \FOUNDcmstdr{} & \FOUNDcCmmean{} & \FOUNDcCmstdl{} & \FOUNDcCmstdr{} \\
10.6 & \FOUNDcnmean{} & \FOUNDcnstdl{} & \FOUNDcnstdr{} & \FOUNDcCnmean{} & \FOUNDcCnstdl{} & \FOUNDcCnstdr{} \\
10.8 & \FOUNDcomean{} & \FOUNDcostdl{} & \FOUNDcostdr{} & \FOUNDcComean{} & \FOUNDcCostdl{} & \FOUNDcCostdr{} \\
11.0 & \FOUNDcpmean{} & \FOUNDcpstdl{} & \FOUNDcpstdr{} & \FOUNDcCpmean{} & \FOUNDcCpstdl{} & \FOUNDcCpstdr{} 
    \end{tabular}
    \caption{Foundation parent colour populations}
    \label{tab:FOUND_parents_c}
\end{table}

\begin{table}
    \centering
    \begin{tabular}{ccccccc}
        Mass & Weight (G10) & $\overline{x}_1$ (G10) & $\sigma$ (G10) & Weight (C11) & $\overline{x}_1$ (C11) & $\sigma$ (C11) \\ 
        \hline
10.0 & \FOUNDxkweightone{} & \FOUNDxkmeanone{} & \FOUNDxkstdone{} & \FOUNDxCkweighttwo{} & \FOUNDxCkmeantwo{} & \FOUNDxCkstdtwo{} \\
10.2 & \FOUNDxlweightone{} & \FOUNDxlmeanone{} & \FOUNDxlstdone{} & \FOUNDxClweighttwo{} & \FOUNDxClmeantwo{} & \FOUNDxClstdtwo{} \\
10.4 & \FOUNDxmweightone{} & \FOUNDxmmeanone{} & \FOUNDxmstdone{} & \FOUNDxCmweighttwo{} & \FOUNDxCmmeantwo{} & \FOUNDxCmstdtwo{} \\
10.6 & \FOUNDxnweightone{} & \FOUNDxnmeanone{} & \FOUNDxnstdone{} & \FOUNDxCnweighttwo{} & \FOUNDxCnmeantwo{} & \FOUNDxCnstdtwo{} \\
10.8 & \FOUNDxoweightone{} & \FOUNDxomeanone{} & \FOUNDxostdone{} & \FOUNDxCoweighttwo{} & \FOUNDxComeantwo{} & \FOUNDxCostdtwo{} \\
11.0 & \FOUNDxpweightone{} & \FOUNDxpmeanone{} & \FOUNDxpstdone{} & \FOUNDxCpweighttwo{} & \FOUNDxCpmeantwo{} & \FOUNDxCpstdtwo{} 
    \end{tabular}
    \caption{Foundation parent stretch populations}
    \label{tab:FOUND_parents_x1}
\end{table}

\begin{table}
    \centering
    \begin{tabular}{ccccccc}
        Mass & mean (G10) & $\sigma_-$ (G10) & $\sigma_+$ (G10) & mean (C11) & $\sigma_-$ (C11) & $\sigma_+$ (C11) \\ 
        \hline
8.2 & \HIZcbmean{} & \HIZcbstdl{} & \HIZcbstdr{} & \HIZcCbmean{} & \HIZcCbstdl{} & \HIZcCbstdr{} \\
8.4 & \HIZccmean{} & \HIZccstdl{} & \HIZccstdr{} & \HIZcCcmean{} & \HIZcCcstdl{} & \HIZcCcstdr{} \\
8.6 & \HIZcdmean{} & \HIZcdstdl{} & \HIZcdstdr{} & \HIZcCdmean{} & \HIZcCdstdl{} & \HIZcCdstdr{} \\
8.8 & \HIZcemean{} & \HIZcestdl{} & \HIZcestdr{} & \HIZcCemean{} & \HIZcCestdl{} & \HIZcCestdr{} \\
9.0 & \HIZcfmean{} & \HIZcfstdl{} & \HIZcfstdr{} & \HIZcCfmean{} & \HIZcCfstdl{} & \HIZcCfstdr{} \\
9.2 & \HIZcgmean{} & \HIZcgstdl{} & \HIZcgstdr{} & \HIZcCgmean{} & \HIZcCgstdl{} & \HIZcCgstdr{} \\
9.4 & \HIZchmean{} & \HIZchstdl{} & \HIZchstdr{} & \HIZcChmean{} & \HIZcChstdl{} & \HIZcChstdr{} \\
9.6 & \HIZcimean{} & \HIZcistdl{} & \HIZcistdr{} & \HIZcCimean{} & \HIZcCistdl{} & \HIZcCistdr{} \\
9.8 & \HIZcjmean{} & \HIZcjstdl{} & \HIZcjstdr{} & \HIZcCjmean{} & \HIZcCjstdl{} & \HIZcCjstdr{} \\
10.0 & \HIZckmean{} & \HIZckstdl{} & \HIZckstdr{} & \HIZcCkmean{} & \HIZcCkstdl{} & \HIZcCkstdr{} \\
10.2 & \HIZclmean{} & \HIZclstdl{} & \HIZclstdr{} & \HIZcClmean{} & \HIZcClstdl{} & \HIZcClstdr{} \\
10.4 & \HIZcmmean{} & \HIZcmstdl{} & \HIZcmstdr{} & \HIZcCmmean{} & \HIZcCmstdl{} & \HIZcCmstdr{} \\
10.6 & \HIZcnmean{} & \HIZcnstdl{} & \HIZcnstdr{} & \HIZcCnmean{} & \HIZcCnstdl{} & \HIZcCnstdr{} \\
10.8 & \HIZcomean{} & \HIZcostdl{} & \HIZcostdr{} & \HIZcComean{} & \HIZcCostdl{} & \HIZcCostdr{} \\
11.0 & \HIZcpmean{} & \HIZcpstdl{} & \HIZcpstdr{} & \HIZcCpmean{} & \HIZcCpstdl{} & \HIZcCpstdr{} \\
11.2 & \HIZcqmean{} & \HIZcqstdl{} & \HIZcqstdr{} & \HIZcCqmean{} & \HIZcCqstdl{} & \HIZcCqstdr{} \\
11.4 & \HIZcrmean{} & \HIZcrstdl{} & \HIZcrstdr{} & \HIZcCrmean{} & \HIZcCrstdl{} & \HIZcCrstdr{} \\
11.6 & \HIZcsmean{} & \HIZcsstdl{} & \HIZcsstdr{} & \HIZcCsmean{} & \HIZcCsstdl{} & \HIZcCsstdr{} \\
11.8 & \HIZctmean{} & \HIZctstdl{} & \HIZctstdr{} & \HIZcCtmean{} & \HIZcCtstdl{} & \HIZcCtstdr{} \\
12.0 & \HIZcumean{} & \HIZcustdl{} & \HIZcustdr{} & \HIZcCumean{} & \HIZcCustdl{} & \HIZcCustdr{} 
    \end{tabular}
    \caption{Combined SDSS, DES, SNLS, PS1 parent colour populations}
    \label{tab:COMB_parents_x1}
\end{table}

\begin{table}
    \centering
    \begin{tabular}{ccccccc}
        Mass & mean (G10) & $\sigma_-$ (G10) & $\sigma_+$ (G10) & mean (C11) & $\sigma_-$ (C11) & $\sigma_+$ (C11) \\ 
        \hline
8.2 & \HIZxbmean{} & \HIZxbstdl{} & \HIZxbstdr{} & \HIZxCbmean{} & \HIZxCbstdl{} & \HIZxCbstdr{} \\
8.4 & \HIZxcmean{} & \HIZxcstdl{} & \HIZxcstdr{} & \HIZxCcmean{} & \HIZxCcstdl{} & \HIZxCcstdr{} \\
8.6 & \HIZxdmean{} & \HIZxdstdl{} & \HIZxdstdr{} & \HIZxCdmean{} & \HIZxCdstdl{} & \HIZxCdstdr{} \\
8.8 & \HIZxemean{} & \HIZxestdl{} & \HIZxestdr{} & \HIZxCemean{} & \HIZxCestdl{} & \HIZxCestdr{} \\
9.0 & \HIZxfmean{} & \HIZxfstdl{} & \HIZxfstdr{} & \HIZxCfmean{} & \HIZxCfstdl{} & \HIZxCfstdr{} \\
9.2 & \HIZxgmean{} & \HIZxgstdl{} & \HIZxgstdr{} & \HIZxCgmean{} & \HIZxCgstdl{} & \HIZxCgstdr{} \\
9.4 & \HIZxhmean{} & \HIZxhstdl{} & \HIZxhstdr{} & \HIZxChmean{} & \HIZxChstdl{} & \HIZxChstdr{} \\
9.6 & \HIZximean{} & \HIZxistdl{} & \HIZxistdr{} & \HIZxCimean{} & \HIZxCistdl{} & \HIZxCistdr{} \\
9.8 & \HIZxjmean{} & \HIZxjstdl{} & \HIZxjstdr{} & \HIZxCjmean{} & \HIZxCjstdl{} & \HIZxCjstdr{} \\
10.0 & \HIZxkmean{} & \HIZxkstdl{} & \HIZxkstdr{} & \HIZxCkmean{} & \HIZxCkstdl{} & \HIZxCkstdr{} \\
10.2 & \HIZxlmean{} & \HIZxlstdl{} & \HIZxlstdr{} & \HIZxClmean{} & \HIZxClstdl{} & \HIZxClstdr{} \\
10.4 & \HIZxmmean{} & \HIZxmstdl{} & \HIZxmstdr{} & \HIZxCmmean{} & \HIZxCmstdl{} & \HIZxCmstdr{} \\
10.6 & \HIZxnmean{} & \HIZxnstdl{} & \HIZxnstdr{} & \HIZxCnmean{} & \HIZxCnstdl{} & \HIZxCnstdr{} \\
10.8 & \HIZxomean{} & \HIZxostdl{} & \HIZxostdr{} & \HIZxComean{} & \HIZxCostdl{} & \HIZxCostdr{} \\
11.0 & \HIZxpmean{} & \HIZxpstdl{} & \HIZxpstdr{} & \HIZxCpmean{} & \HIZxCpstdl{} & \HIZxCpstdr{} \\
11.2 & \HIZxqmean{} & \HIZxqstdl{} & \HIZxqstdr{} & \HIZxCqmean{} & \HIZxCqstdl{} & \HIZxCqstdr{} \\
11.4 & \HIZxrmean{} & \HIZxrstdl{} & \HIZxrstdr{} & \HIZxCrmean{} & \HIZxCrstdl{} & \HIZxCrstdr{} \\
11.6 & \HIZxsmean{} & \HIZxsstdl{} & \HIZxsstdr{} & \HIZxCsmean{} & \HIZxCsstdl{} & \HIZxCsstdr{} \\
11.8 & \HIZxtmean{} & \HIZxtstdl{} & \HIZxtstdr{} & \HIZxCtmean{} & \HIZxCtstdl{} & \HIZxCtstdr{} \\
12.0 & \HIZxumean{} & \HIZxustdl{} & \HIZxustdr{} & \HIZxCumean{} & \HIZxCustdl{} & \HIZxCustdr{} 
    \end{tabular}
    \caption{Combined SDSS, DES, SNLS, PS1 parent stretch populations}
    \label{tab:COMB_parents_c}
\end{table}

\clearpage

\bibliographystyle{mne2.bst}
\bibliography{research2.bib}
\end{document}